\documentclass{article}

\usepackage{microtype}
\usepackage{graphicx}
\usepackage{booktabs} 
\usepackage{xspace}
\usepackage{algorithm}
\usepackage{algorithmic}
\usepackage{hyperref}
\usepackage{multirow}
\usepackage{pifont}
\usepackage[inline]{enumitem}
\usepackage{times}
\usepackage{latexsym}
\usepackage{gensymb}
\usepackage{textcomp}
\usepackage{subcaption}
\usepackage{colortbl}
\usepackage{makecell}

\newcommand{\RETURN}{\STATE \textbf{return} }
\usepackage[table,xcdraw]{xcolor}

\usepackage[accepted]{icml2025}

\usepackage{amsmath}
\usepackage{amssymb}
\usepackage{mathtools}
\usepackage{amsthm}

\usepackage[capitalize,noabbrev]{cleveref}

\theoremstyle{plain}

\theoremstyle{definition}

\theoremstyle{remark}

\usepackage[textsize=tiny]{todonotes}

\icmltitlerunning{KGMark: A Diffusion Watermark for Knowledge Graphs}

\begin{document}

\newcommand{\sysname}{{{KGMark}}\xspace}

\twocolumn[
\icmltitle{KGMark: A Diffusion Watermark for Knowledge Graphs}

\icmlsetsymbol{equal}{*}

\begin{icmlauthorlist}
\icmlauthor{Hongrui Peng}{equal,bupt}
\icmlauthor{Haolang Lu}{equal,bupt}
\icmlauthor{Yuanlong Yu}{bupt}
\icmlauthor{Weiye Fu}{bupt}
\icmlauthor{Kun Wang}{ntu}
\icmlauthor{Guoshun Nan}{bupt}

\end{icmlauthorlist}


\icmlaffiliation{bupt}{Beijing University Of Posts and Telecommunications, Beijing, China}
\icmlaffiliation{ntu}{Nanyang Technological University, Singapore}

\icmlcorrespondingauthor{Guoshun Nan}{nanguo2021@bupt.edu.cn}
\icmlcorrespondingauthor{Kun Wang}{wang.kun@ntu.edu.sg}

\icmlkeywords{Watermarking, Synthetic Graph, Generative Model, Multi-Modal}

\vskip 0.3in
]

\printAffiliationsAndNotice{\icmlEqualContribution} 

\begin{abstract}

Knowledge graphs (KGs) are ubiquitous in numerous real-world applications, and watermarking facilitates protecting intellectual property and preventing potential harm from AI-generated content. Existing watermarking methods mainly focus on static plain text or image data, while they can hardly be applied to dynamic graphs due to spatial and temporal variations of structured data. This motivates us to propose \sysname, the first graph watermarking framework that aims to generate robust, detectable, and transparent diffusion fingerprints for dynamic KG data. Specifically, we propose a novel clustering-based alignment method to adapt the watermark to spatial variations. Meanwhile, we present a redundant embedding strategy to harden the diffusion watermark against various attacks, facilitating the robustness of the watermark to the temporal variations. Additionally, we introduce a novel learnable mask matrix to improve the transparency of diffusion fingerprints. By doing so, our \sysname properly tackles the variation challenges of structured data. Experiments on various public benchmarks show the effectiveness of our proposed \sysname.

\end{abstract}
\vspace{-2em}
\section{Introduction}

The growing adoption of generative models has significantly expanded the creation and utilization of synthetic data~\cite{bauer2024comprehensiveexplorationsyntheticdata}, including structured formats such as time-series data~\cite{10.5555/3692070.3692474}, tabular data~\cite{10.5555/3692070.3694089}, and graphs~\cite{han2025retrievalaugmentedgenerationgraphsgraphrag, wang2023brave}.
Among these, Knowledge graphs (KGs)~\cite{Ji_2022} are especially crucial due to their ability to represent complex relationships and semantic hierarchies~\cite{10.5555/3692070.3692129}, making them indispensable for applications such as semantic search~\cite{10.5555/3666122.3667344}, question answering~\cite{yin2024rethinking,zhou2024less}, and recommendation systems~\cite{10.1145/3298689.3347011,10.1145/3331184.3331267}.
Deep learning-based models such as GraphRNN~\cite{pmlr-v80-you18a}, GraphVAE~\cite{10.1007/978-3-030-01418-6_41}, MolGAN~\cite{decao2022molganimplicitgenerativemodel}, and DiGress~\cite{vignac2023digressdiscretedenoisingdiffusion} have emerged as powerful tools for generating high-quality synthetic graphs, supporting applications such as graph classification, molecular design, and structure-preserving data augmentation in machine learning pipelines.
\begin{figure}[t]
\centering
\includegraphics[width=0.48\textwidth]{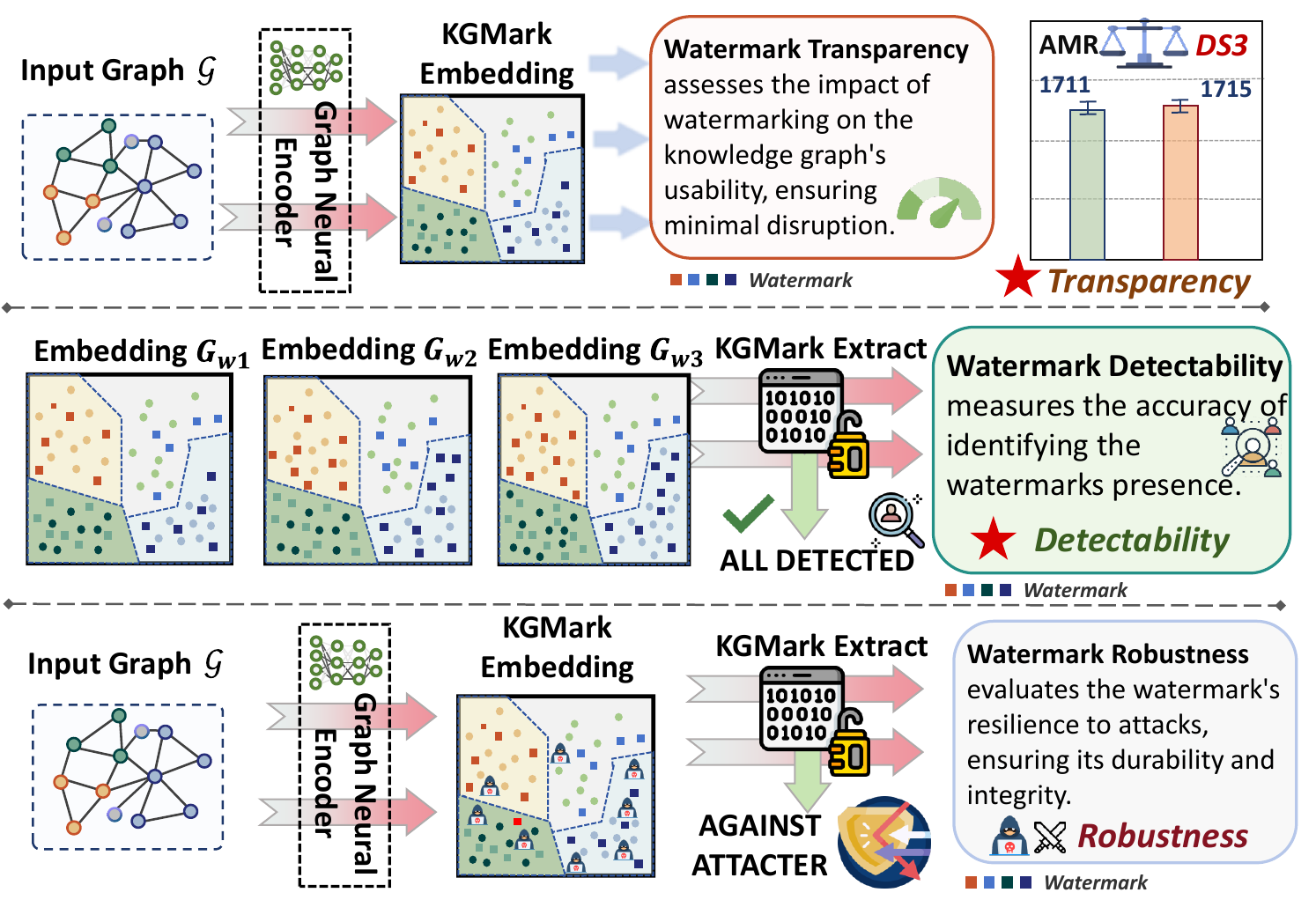}
\vspace{-2em}
\caption{Overview of our \sysname. \sysname implements a KGE watermarking scheme that preserves transparency, enables reliable detection, and remains robust against various post-editing attacks.} 
\label{fig: intro}
\vspace{-5mm}
\end{figure}

However, existing synthetic graph generation methods may inadvertently embed biases~\cite{10.1145/3543507.3583544,10.1145/3534678.3539319} or misleading~\cite{Yang2024} information and are vulnerable to malicious alterations by attackers~\cite{ijcai2019p674}, potentially introducing harmful content that compromises analyses or even facilitates the exploitation of real-world systems~\cite{jiang2024kgfit,wang2025comprehensive}.
Furthermore, presenting synthetic graphs as original research can violate intellectual property rights, undermining trust in academic and commercial environments~\cite{Smits2022, Zhu2024WatermarkembeddedAE}.
Ensuring \textit{traceability}, \textit{integrity}, and \textit{copyright} protection of synthetic counterparts causes a significant amount of compute and memory footprints due to the notorious designs~\cite{liang2024clustering,10.24963/ijcai.2024/364}, which even unnerves large corporations, let alone individual researchers.

Watermarking techniques~\cite{DBLP:conf/icml/RadfordKHRGASAM21, podell2023sdxlimprovinglatentdiffusion} have proven to be effective in ensuring the authenticity of synthetic data in image and text generation~\cite{NEURIPS2023_b54d1757, zhao2023provablerobustwatermarkingaigenerated}.
However, off-the-shelf watermarking algorithms~\cite{liu2024surveytextwatermarkingera} appear to lag behind in the era of graph structures, despite their increasing importance in various domains.
This gap makes watermarking for KGs an unsolved and looming challenge.

The core challenges of graph watermarking lie in achieving robustness against spatial-temporal variations and maintaining transparency.
Embedded watermarks must preserve KG semantics while resisting structural perturbations caused by dynamic node/edge updates (e.g., temporal evolution in recommendation systems~\cite{10.1145/3331184.3331267}) and graph isomorphism (e.g., node relabeling or spatial rearrangements~\cite{9721082}). 
Conventional watermarking methods~\cite{9721082}, designed for static and unstructured data, often exhibit limited robustness under the spatial-temporal variations of dynamic KGs, posing challenges in mitigating risks of AI-generated KG misuse~\cite{10.24963/ijcai.2024/243}.
The inherent heterogeneity of knowledge graphs further necessitates embedding watermarks at the knowledge graph embedding (KGE) level to reconcile semantic fidelity with detection resilience~\cite{pmlr-v235-le24c}.

To address these challenges, we propose \sysname, the first watermarking scheme specifically designed for knowledge graphs, based on diffusion models.
 \sysname employs graph alignment with a learnable mask matrix to spatially adapt watermarks to structural variations, ensuring seamless integration between topology and embedded signals.
For temporal robustness, we design redundant embedding strategies coupled with likelihood estimation, enabling resilient extraction despite incremental graph updates. 
\sysname requires no prior assumptions about graph stability or isomorphism, making it agnostic to node ordering and scalable to evolving KGs.
The watermark is detectable exclusively through secure keys, ensuring resistance to adversarial attacks while preserving KG utility~\cite{Nandi2023DynaSembleDE}.
 
Through rigorous testing, we demonstrate that \sysname achieves: \ding{182} high detectability, with a watermark detection AUC up to $0.99$; \ding{183} maintaining KG quality and limiting downstream task performance loss to within the range of $0.02\% \sim 9.7\%$; and \ding{184} high robustness, retaining an AUC of around $0.95$ against various post-editing attacks.

Briefly, our key contributions are summarized as follows:
\vspace{-1em}
\begin{itemize}[leftmargin=*]
    \item[\ding{224}]We introduce \sysname, \textit{the first watermarking scheme designed for KGE,} embedding robust and transparent watermarks into graph structures to protect intellectual property and ensure data integrity.
    \vspace{-0.5em}
    \item[\ding{224}]We propose effective solutions to key challenges, including redundancy-based embedding and the Learnable Adaptive Watermark Matrix, which significantly enhance the performance of \sysname.
    \vspace{-0.5em}
    \item[\ding{224}]We demonstrate through rigorous experiments the superior performance of \sysname, showing its robustness, transparency, and high detectability across various post-editing attack scenarios.
\end{itemize}


\vspace{-0.8em}
\section{Related Work}

\vspace{-0.3em}
\subsection{Watermarking AI-generated Content}
\vspace{-0.4em}

As AI technologies continue to evolve and generate increasingly sophisticated content, the necessity for robust watermarking solutions for AI-generated content has gained prominence~\cite{zhao2023invisible, zhu2024genimage}. This technique serves a critical role in addressing issues such as copyright infringement~\cite{zhang2024editguard}, misinformation, and the ethical implications surrounding the use of AI in creative processes. In particular, watermarking helps to distinguish AI-generated work from human-created content~\cite{asnani2024promark}, thereby enhancing credibility in the digital landscape~\cite{barman2024brittleness}.

\vspace{-0.3em}
\subsection{Watermarking Synthetic Unstructured Data} 
\vspace{-0.4em}

Watermarking synthetic unstructured data, such as plain text~\cite{Dathathri2024} and images~\cite{zhang2024attackresilient, wen2024tree}, is crucial for ensuring data integrity, protecting intellectual property, and preventing misuse. Common methods include pre-applying watermarks to training datasets, enabling models to generate inherently watermarked data~\cite{yu2021artificial, zhao2023recipe}, or modifying the sampling process in large language models to subtly alter output word distributions and encode watermarks~\cite{kirchenbauer2023watermark}. A widely used approach in diffusion models involves embedding watermarks into the initial noise vector or latent space without retraining~\cite{wen2024tree, Yang_2024_CVPR, yang2024guisegraphgaussianshading}, ensuring transparency and robustness through reversibility. In this work, we conduct the first systematic exploration of adaptively incorporating this method into knowledge graphs.

\begin{figure*}
\centering
\includegraphics[width=1.0\textwidth]{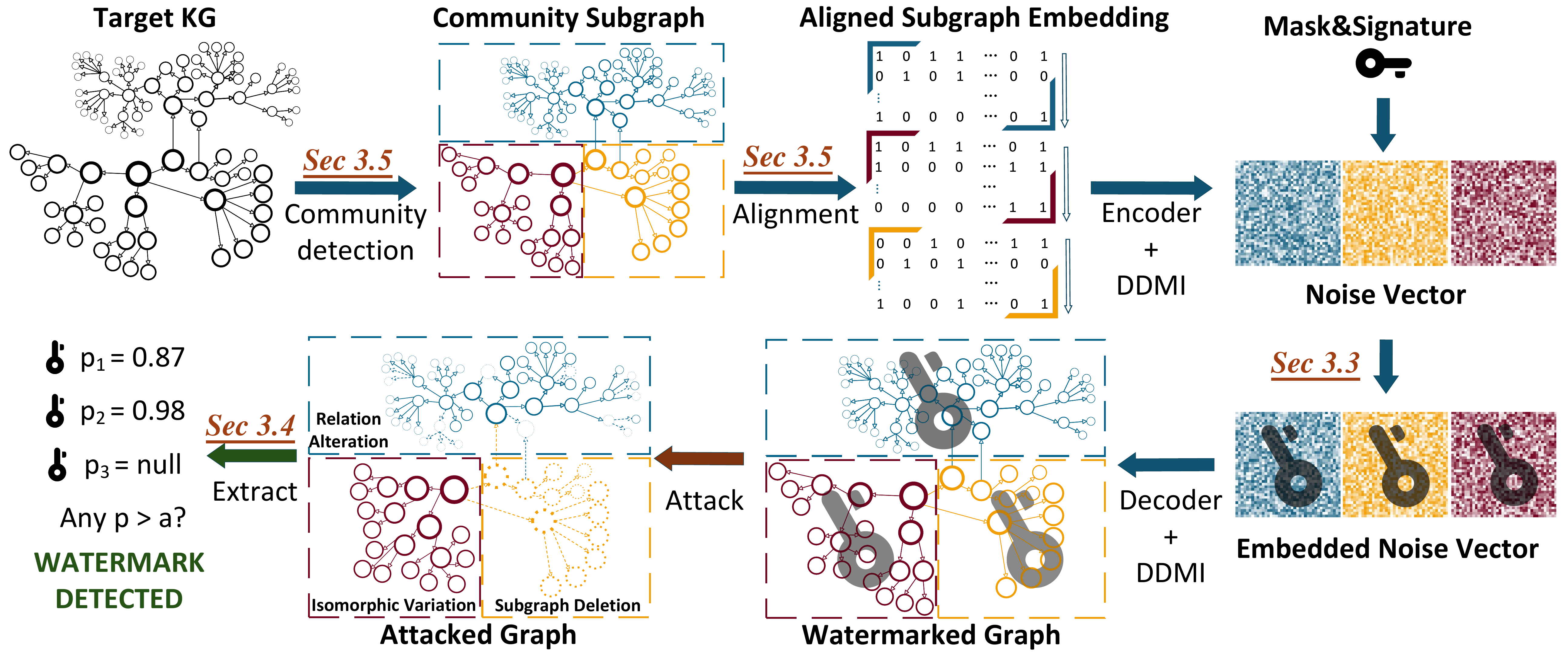}
\caption{Pipeline of the proposed \sysname. The target KGE undergoes community detection and alignment before watermark embedding, enabling robust watermark extraction under attacks (Relation Alteration, Subgraph Deletion, and Isomorphic Variation).} 
\label{fig: System Overview}
\vspace{-0.3cm}
\end{figure*}

\vspace{-0.5em}
\section{Proposed Method: \sysname}
\vspace{-0.3em}
\subsection{Preliminary of Latent Diffusion Model}
\vspace{-0.4em}

The core idea of the Diffusion Model can be summarized in two phases: the forward diffusion process and the reverse diffusion process~\cite{chang2023designfundamentalsdiffusionmodels, rombach2021highresolution}. In the forward diffusion process, the latent representation $\mathcal{Z}_0$ is progressively corrupted by Gaussian noise over $\mathcal{T}$ steps, resulting in:
\begin{equation}
q(\mathcal{Z}_t | \mathcal{Z}_{t-1}) = \mathcal{N}(\mathcal{Z}_t; \sqrt{1 - \beta_t} \mathcal{Z}_{t-1}, \beta_t I)
\end{equation}
As $t \to \mathcal{T}$, $\mathcal{Z}_\mathcal{T}$ converges to an approximate \textit{standard Gaussian distribution}.
The reverse diffusion process starting from noise $\mathcal{Z}_\mathcal{T}$, the model denoises it step-by-step to recover $\mathcal{Z}_0$:
\begin{equation}
p_\theta(\mathcal{Z}_{t-1} | \mathcal{Z}_t) = \mathcal{N}(\mathcal{Z}_{t-1}; \mu_\theta(\mathcal{Z}_t, t), \Sigma_\theta(\mathcal{Z}_t, t))
\end{equation}

Our method employs a latent diffusion model (LDM) with DDIM~\cite{song2022denoisingdiffusionimplicitmodels} for sampling. DDIM's deterministic inversion process allows efficient recovery of the initial noise vector $\mathcal{Z}_\mathcal{T}^{\text{INV}}$, which is then used for watermark embedding, ensuring imperceptibility and integrity preservation in the latent space. The precise control offered by DDIM's reverse diffusion process is key to balancing watermark robustness and reconstruction fidelity.

\vspace{-0.3em}
\subsection{Overview of \sysname}
\vspace{-0.3em}
The proposed watermarking framework embeds a watermark into a KGE $\mathcal{G} = (\mathcal{V}, \mathcal{E})$. 
The process begins by encoding $\mathcal{G}$ into a latent representation $\mathcal{Z}_0$ using a graph encoder, trained under a VAE framework. 

\textbf{Watermark Embedding:} A signature $\mathcal{S}$ is embedded into designated latent subspaces:
\begin{equation}
        \Delta = \mathbf{F}(\mathcal{Z}_{\mathcal{T}}^{\text{INV}}) \cdot (1 - \mathbb{M}) + \mathbf{F}(\mathcal{S}) \cdot \mathbb{M}, \mathbb{M} \in \{0, 1\}^{m \times n},
    \label{equ: delta}
\end{equation}
where $\mathbf{F}(\cdot)$ denotes the Fourier transform, and $\mathbb{M}$ is a mask matrix defining embedding regions for $\mathcal{S}$. The watermarked latent vector is then derived via inverse Fourier transform:
\begin{equation}
    \mathcal{Z}_{\mathcal{T}}^w = \mathbf{F}^{-1}(\Delta), \quad \mathcal{S} \sim \mathcal{N}(0, \sigma^2 \mathbf{I}),
    \label{equ: zt_w}
\end{equation}
Where $\sigma^2$ determines the variance of the \textit{normal distribution}. $\mathcal{Z}_{\mathcal{T}}^w$ undergoes reverse diffusion, yielding $\mathcal{Z}_0^w$, which is decoded to reconstruct the watermarked graph $\mathcal{G}^w$.

\textbf{Attack Modeling:}
We define the attack intensity by the total perturbation $\delta$ applied to the graph $\mathcal{G}$, where $\delta = \sum_{k} \delta_k$, and each $\delta_k$ denotes the perturbation (e.g., post-editing operations) applied to a subgraph $\mathcal{G}_{sub}$.

The adversary aims to invalidate the watermark by reducing the similarity between the extracted watermark and the original signature $\mathcal{S}$. 
Let $T(\cdot)$ denote the watermark extraction function. The attack objective is then formulated as:
\vspace{-0.4em}
\begin{equation}
\min_{\delta} \, \text{sim}(T(\mathcal{G}^w + \delta), \mathcal{S}) + \gamma \cdot \sum_{k} \|\delta_k\|_q,
\label{equ: attack}
\vspace{-0.4em}
\end{equation}
where $\text{sim}(\cdot, \cdot)$ denotes a similarity measure, and $\gamma$ is a trade-off parameter that controls the allowed perturbation budget. A successful attack reduces the similarity score while maintaining the semantic validity of subgraphs.

As illustrated in \Cref{fig: System Overview}, isomorphic variations preserve local subgraph structure, while stronger perturbations compromise the global topology. This highlights another unique characteristic of graph-based watermarking: \textit{the adversary can strategically degrade global structural fidelity while preserving local usability}.

\textbf{Watermark Extraction:} For watermark extraction, $\mathcal{G}^w$ is encoded and inverted to approximate $\mathcal{Z}_{\mathcal{T}}^{\text{INV}}$. The Fourier transform of $\mathcal{Z}_{\mathcal{T}}^{\text{INV}}$ is computed as:
\begin{equation}
    \mathcal{Y} = \mathbf{F}(\mathbf{Z}_\mathcal{T}^\text{INV}), \quad \text{s.t.} \quad \mathbb{E}[\mathcal{Y}_i] = \mu_i, \ \text{Var}[\mathcal{Y}_i] = \sigma_i^2,
    \label{equ: noise}
\end{equation}
A test statistic $\mathcal{T}$ is computed by comparing $\mathcal{Y}$ with $(\mathcal{S}, \mathbb{M})$. The P-value is evaluated using a noncentral $\chi^2$ distribution, and the watermark is detected if the P-value is below a predefined significance level.

To fulfill the three essential properties of a watermarking \textbf{transparency} during embedding, \textbf{robustness} against adversarial perturbations, and \textbf{detectability} upon extraction, \sysname introduces three dedicated designs across the embedding, extraction, and attack resilience stages. 

Specifically, we present: 
(1) a \textit{Learnable Adaptive Watermark Mask Matrix} (Sec.~\ref{method: Learnable Adaptive Watermark Mask Matrix}) to ensure transparent watermark embedding, 
(2) a \textit{Defending Against Isomorphism and Structural Variations} module (Sec.~\ref{method: Defending Against Isomorphism and Structural Variations}) to enhance watermark robustness under perturbations, and 
(3) a \textit{Likelihood-Based Watermark Verification} mechanism (Sec.~\ref{method: Likelihood-Based Watermark Verification}) validate the embedded watermark.

\vspace{-0.4em}
\subsection{Learnable Adaptive Watermark Mask Matrix} \label{method: Learnable Adaptive Watermark Mask Matrix}
\vspace{-0.4em}

\textit{Principle 3.1 (\textbf{Latent Space Equilibrium}).}
\textit{To preserve the latent space equilibrium during watermark embedding, we decompose the total perturbation into two parts: (1) the inherent noise from DDIM inversion, and (2) the additional distortion from watermark embedding. Let $\mathcal{Z}_{\mathcal{T}-k_j}^{\text{INV}}$ be the $k_j$ step latent from DDIM inversion, $\widehat{\mathcal{Z}}_{\mathcal{T}-k_j}$ the latent from DDIM without watermark, and $\mathcal{Z}_{\mathcal{T}-k_j}^w$ the latent with watermark. The total deviation satisfies:}
\vspace{-0.5em}
\begin{equation}
\sum_{j=1}^{\mathcal{T}} \underbrace{\left\| \widehat{\mathcal{Z}}_{\mathcal{T}-k_j} - \mathcal{Z}_{\mathcal{T}-k_j}^{\text{INV}} \right\|^2}_{\text{DDIM sampling loss}} + 
\underbrace{\left\| \mathcal{Z}^w_{\mathcal{T}-k_j} - \widehat{\mathcal{Z}}_{\mathcal{T}-k_j} \right\|^2}_{\text{Watermark embedding loss}} \leq \epsilon.
\label{equ:principle_decomposed}
\end{equation}
\textit{This formulation isolates the unavoidable sampling error from the controllable embedding distortion, guiding watermark design to minimize overall latent drift.}

To operationalize \textit{\textbf{Latent Space Equilibrium}}, we design a learnable masking mechanism that controls how watermark signals are injected into the latent space.
Watermark embedding inevitably introduces reconstruction loss $\mathcal{L}$ for $\mathcal{G}$, quantified by the discrepancy between the $\mathcal{Z}_\mathcal{T}^{\text{INV}}$ and $\mathcal{Z}_\mathcal{T}^{w}$~\cite{fridrich2002lossless}.
Specifically, we introduce a learnable, adaptive mask matrix $\mathbb{M}$,  optimized for the structure of $\mathcal{G}$. 
$\mathbb{M}$ is trained to minimize $\mathcal{L}$ in Equation~\eqref{equ: L1}, which quantifies the discrepancy between the $\mathcal{Z}_\mathcal{T}^{\text{INV}}$ and $\mathcal{Z}_\mathcal{T}^{w}$. 

\begin{equation}
\mathcal{L} = \sum_{j \in [1, \mathcal{T}]} \left\| \mathcal{Z}_{\mathcal{T}-k_j}^{\text{INV}} - f_{\text{DDIM}}^{k_j} \left( f_w(\mathcal{Z}_\mathcal{T}^{\text{INV}}, \mathcal{S}, \mathbb{M}), \mathcal{T} \right) \right\|^2,
\label{equ: L1}
\end{equation}
where $\mathcal{T}$ denotes the total diffusion steps, and $k_j$ satisfies $0 < k_j < \mathcal{T}$. The function $f_w$ represents the watermark embedding function, while $f_{\text{DDIM}}^{k_j}$ denotes the DDIM sampling process at time step $k_j$.

To further refining, we adopt a "sample-then-embed" strategy with a correction term $\alpha \mathcal{S} \cdot \mathbb{M}$ ($\alpha$ is a tunable coefficient), ensuring better alignment in the latent space.
\begin{equation}
\begin{split}
\mathcal{L} &= \sum_{j \in [1, \mathcal{T}]} \left\| \mathcal{Z}_{\mathcal{T}-k_j}^{\text{INV}} \right. - \\
&\quad \left. \left[ f_w\left(f_{\text{DDIM}}^{k_j}(\mathcal{Z}_\mathcal{T}^{\text{INV}}, \mathcal{T}), \mathcal{S}, \mathbb{M}\right) + \alpha \mathcal{S} \cdot \mathbb{M} \right] \right\|^2.
\end{split}
\label{equ: L2}
\end{equation}
As $\mathcal{L}$ shown in Equation~\eqref{equ: L2}, $\mathcal{S} \sim \mathcal{N}(0, \mathbf{I})$ represents a noise vector sampled from a standard normal distribution, where $\mathbf{I}$ is the identity matrix. The adaptive nature of $\mathbb{M}$ ensures that it is specific to each graph $\mathcal{G}$, making the watermark embedding robust to structural variations.

To characterize the sparsity of the learnable mask matrix $\mathbb{M}$, we define its density as:
\begin{equation} 
Density(\mathbb{M}) = \frac{\sum_{i=1}^m\sum_{j=1}^n\mathbb{M}_{ij}}{m \times n}, \label{equ: density} \end{equation}
where $m \times n$ is the total number of elements in $\mathbb{M}$. This metric quantifies the proportion of nonzero entries, balancing watermark imperceptibility and robustness against attacks.

During watermark extraction, we evaluate all candidate $\mathbb{M}$. The decision rule is formulated as Equation~\eqref{equ: decision}: 
\begin{equation}
\mathcal{G}_{\text{classified}} = 
\begin{cases} 
\top, & \exists \mathbb{M} \, \text{s.t.} \, d\big(f_{\text{ex}}(\mathbb{M}, \mathcal{G}), \mathcal{S}\big) \leq \delta, \\
\bot, & \forall \mathbb{M}, \, d\big(f_{\text{ex}}(\mathbb{M}, \mathcal{G}), \mathcal{S}\big) > \delta,
\end{cases}
\label{equ: decision}
\end{equation}
where $f_{\text{ex}}(\mathbb{M}, \mathcal{G})$ denotes the extracted signature from $\mathcal{G}$ using matrix $\mathbb{M}$, $d(\cdot, \cdot)$ is the distance metric quantifying the difference between signatures, and $\delta$ is the threshold determining signature validity.

\vspace{-0.4em}
\subsection{Likelihood-Based Watermark Verification}\label{method: Likelihood-Based Watermark Verification}
\vspace{-0.4em}

In \sysname, watermark detection utilizes a likelihood ratio test, where the null hypothesis $\mathcal{H}_0$ assumes the noise vector $\mathcal{Y}$ follows a Gaussian distribution $\mathcal{N}(\mathbf{0}, \sigma^2\mathcal{I}_\mathbb{C})$. Under the alternative hypothesis $\mathcal{H}_1$, the graph contains a watermark, introducing a detectable deviation from this distribution.

To integrate the distance metric $d(\cdot, \cdot)$ and the signature extraction function $f_{\text{ex}}$, we define the residual vector $\mathcal{R}$ as the difference between the extracted signature and the optimal reference signature $\mathcal{K}^* \in \mathcal{S}$:
\begin{equation}
\mathcal{R} = f_{\text{ex}}(\mathbb{M}, \mathcal{G}) - \mathcal{K}^*.
\label{equ: opti ref}
\end{equation}
The distance metric $d(\cdot, \cdot)$ thus measures the magnitude of $\mathcal{R}$, which quantifies the deviation between the extracted signature and the expected reference signature.
The likelihood of observing $\mathcal{R}$ under $\mathcal{H}_0$ is given by Equation~\eqref{equ: likelihood}:
\begin{equation}
\mathcal{L}(\mathcal{R} | \mathcal{H}_0) = \prod_{i \in \mathbb{M}} \frac{1}{\sqrt{2\pi \sigma^2}} \exp\left(-\frac{|\mathcal{R}_i|^2}{2\sigma^2}\right),
\label{equ: likelihood}
\end{equation}
where $\sigma^2 = \frac{1}{|\mathbb{M}|} \sum_{i \in \mathbb{M}} |\mathcal{R}_i|^2$ is the estimated variance of the residual vector, computed as the mean squared deviation of the residuals within the mask region 
$\mathbb{M}$.

The likelihood ratio test statistic $\lambda$ is defined as the logarithmic ratio of the likelihoods under $\mathcal{H}_0$ and $\mathcal{H}_1$:
\begin{equation}
\lambda = -2 \log \frac{\sup_{\mathcal{H}_0} \mathcal{L}(\mathcal{R} | \mathcal{H}_0)}{\sup_{\mathcal{H}_1} \mathcal{L}(\mathcal{R} | \mathcal{H}_1)}.
\label{equ: likelihood_ratio}
\end{equation}
For watermark detection, we employ a simplified test statistic $\hat{\mathcal{T}}$, defined in Equation~\eqref{equ: test_statistic}, which directly quantifies the deviation of $\mathcal{R}$ from the expected values under $\mathcal{H}_0$:
\begin{equation}
\hat{\mathcal{T}} = \frac{1}{\sigma^2} \sum_{i \in \mathbb{M}} |\mathcal{R}_i|^2.
\label{equ: test_statistic}
\end{equation}
Under $\mathcal{H}_0$, $\hat{\mathcal{T}}$ follows a noncentral chi-squared distribution with degrees of freedom $|\mathbb{M}|$ and noncentrality parameter $\lambda = \frac{1}{\sigma^2} \sum_{i} |\mathcal{K}_i^*|^2$~\cite{patnaik1949non}. The p-value for watermark detection is then calculated using the cumulative distribution function (CDF) of this distribution~\cite{glasserman2004monte}, as shown in \Cref{equ: p_value}:
\begin{equation}
\begin{aligned}
p &= \Pr\left( \chi^2_{|\mathbb{M}|, \lambda} \leq \hat{\mathcal{T}} \mid \mathcal{H}_0 \right) \\
  &= \int_0^{\hat{\mathcal{T}}} \frac{1}{\Gamma\left(\frac{|\mathbb{M}|}{2}\right)} \left(\frac{x}{2}\right)^{\frac{|\mathbb{M}|}{2} - 1} \exp\left( -\frac{x + \lambda}{2} \right) dx.
\end{aligned}
\label{equ: p_value}
\end{equation}
If the $p$ is below a significance level $\alpha$, $\mathcal{G}$ is considered watermarked; otherwise, the null hypothesis $\mathcal{H}_0$ is not rejected.

\vspace{-0.4em}
\subsection{Defending Isomorphism and Structural Variations} \label{method: Defending Against Isomorphism and Structural Variations}
\vspace{-0.4em}

Real-world graph applications encounter isomorphism and structural variations, which are unique to graph-structured data and can be exploited for attacks~\cite{YU2023110534}. Isomorphism refers to graphs that preserve their overall structural connectivity but differ in node ordering or node-specific properties. Structural variations involve more substantial modifications that can alter the graph's topology or the relationship between nodes. See Appendix~\ref{appendix Normalization} for details.

\textbf{Isomorphism Variations:} Graph isomorphism~\cite{yan2002gspan} transformations alter the latent space representation $z_0$ of a graph, despite preserving its inherent properties~\cite{9721082}. To address this, we ensure $z_0$ remains invariant under adjacency matrix reorderings via a \textit{graph alignment} procedure. Let $V(\mathcal{G})$ denote the set of vertices, and $A$ the adjacency matrix. Vertices are reordered by degree $\deg(v_i)$ and clustering coefficient $C(v_i)$:
\begin{equation}
\begin{aligned}
\deg(v_i) > \deg(v_j) \implies i < j, \\
\quad \deg(v_i) = \deg(v_j) \implies C(v_i) \geq C(v_j).
\end{aligned}
\label{equ: imsorp}
\end{equation}
Graph attributes (vertex features, edge weights, adjacency matrix) are adjusted to the new vertex order, preserving structural consistency.

\textbf{Structural Variations:} 
Let $\Delta A \in \{0,1\}^{n \times n}$ denote the perturbation matrix, where $\|\Delta A\|_0 = \delta$ quantifies the attack intensity as the number of edge modifications. 
Structural variations introduce perturbations at multiple scales of graph topology, formalized by the divergence metric:
\begin{equation}
\mathcal{D}(\mathcal{G}, \tilde{\mathcal{G}}) = \underbrace{\|A - \tilde{A}\|_F}_{\text{local perturbations}} + \alpha \underbrace{\|L^\dagger - \tilde{L}^\dagger\|_2}_{\text{global spectral shifts}},
\label{equ: stu var}
\end{equation}
where $L^\dagger$ is the pseudoinverse of the graph Laplacian $L$, and $\alpha$ balances local and global contributions. This divergence aligns with the attacker’s objective in Equation~\eqref{equ: attack}, where the constraint $\|\Delta A\|_0 \leq \delta$ limits the attack intensity, and $sim(T(\mathcal{G}^w + \Delta A), \mathcal{S})$ measures degradation of the extracted watermark.

The local term $\|A - \tilde{A}\|_F$ captures edge-level modifications, including Gaussian noise, Relation Alteration, and Triple Deletion.
The global term $\|L^\dagger - \tilde{L}^\dagger\|_2$ reflects low-frequency structural changes often introduced by smoothing attacks that diffuse information across the graph (Table~\ref{tab: detect robust}).

\textit{Principle 3.2 (\textbf{Information-Theoretic Robustness}).
The watermark embedding $\mathcal{W}$ must ensure that the mutual information $I(\mathcal{S}; T(\mathcal{G}^w + \Delta A))$ between the original watermark $\mathcal{S}$ and the extracted signature under attack satisfies:}  
\begin{equation}
    \inf_{\|\Delta A\|_0 \leq \delta} I(\mathcal{S}; T(\mathcal{G}^w + \Delta A)) \geq \beta,
\end{equation}
\textit{where $\beta > 0$ is a lower bound guaranteeing detectable information retention. This is achieved by enforcing that the watermark encoding $\mathcal{W}(\mathcal{G})$ maximizes the effective information capacity $C(\mathcal{W})$ under adversarial constraints:}
\begin{equation}
C(\mathcal{W}) = \min_{\Delta A} \left[ H(\mathcal{S}) - H(\mathcal{S} \mid T(\mathcal{G}^w + \Delta A)) \right],
\end{equation}
\textit{where $H(\cdot)$ denotes entropy. The principle mandates that critical graph substructures (e.g., high-centrality communities) encode $\mathcal{S}$ with minimal entropy loss $H(\mathcal{S} \mid T(\cdot))$, ensuring robustness against $\mathcal{D}(\mathcal{G}, \tilde{\mathcal{G}})$. }

To satisfy \textit{Principle 3.2}, we partition $\mathcal{G}$ into $l$ communities  \(\{\mathcal{C}_i\}_{i=1}^l\)with vertices ranked by centrality $\eta(v)$. The watermark is embedded as:  
\begin{equation}
\mathcal{W}(\mathcal{G}) = \bigcup_{i=1}^l \Phi(\mathcal{C}_i) \cup \bigcup_{v \in \mathcal{C}_i} \Psi(v),
\end{equation}
where $\Phi(\mathcal{C}_i)$ injects $\mathcal{S}$ into the community’s spectral profile (resilient to $\|L^\dagger - \tilde{L}^\dagger\|_2$), and $\Psi(v)$ encodes $\mathcal{S}$ via edge-weight distributions around high-centrality vertices (resistant to $\|A - \tilde{A}\|_F$). This dual encoding maximizes $C(\mathcal{W})$ by distributing $\mathcal{S}$ across both low-frequency (global) and high-frequency (local) structural invariants.

\begin{table*}[ht]
\centering
\caption{Watermark Detectability \& Robustness (Relation Alteration, Triple Deletion, two adversarial attacks (L2 Metric, NEA), Isomorphism Variation (IsoVar)). We evaluate the detectability of the watermark under clean samples and its robustness under editing attacks with structural and isomorphism variations. Includes four baselines and three variants compared with \sysname.}
\resizebox{\textwidth}{!}{%
\begin{tabular}{l|l|ccccccccccc}
\toprule
\multirow{2}{*}{\textbf{Datasets}} & \multirow{2}{*}{\textbf{Method}} & \multirow{2}{*}{\textbf{Clean}} & \multicolumn{3}{c}{\textbf{Relation Alteration}} & \multicolumn{3}{c}{\textbf{Triple Deletion}} & \multicolumn{2}{c}{\textbf{Adversarial}} & \multirow{2}{*}{\textbf{IsoVar}}  &\multirow{2}{*}{\textbf{Avg}}\\
\cmidrule(lr){4-6} \cmidrule(lr){7-9} \cmidrule(lr){10-11}
& & & \cellcolor{gray!10}10\% & \cellcolor{gray!20}30\% & \cellcolor{gray!30}50\% & \cellcolor{gray!10}10\% & \cellcolor{gray!20}30\% & \cellcolor{gray!30}50\% & L2 Metric & NEA &  \textbf{} & \textbf{} \\
\midrule
\multirow{8}{*}{AliF} 
& \cellcolor{gray!20}DwtDct & \cellcolor{gray!20}0.9837 & \cellcolor{gray!20}0.9730 & \cellcolor{gray!20}0.8813 & \cellcolor{gray!20}0.8371 & \cellcolor{gray!20}0.9457 & \cellcolor{gray!20}0.8577 & \cellcolor{gray!20}0.7724 & \cellcolor{gray!20}0.9577 & \cellcolor{gray!20}0.9638 & \cellcolor{gray!20}0.6039 & \cellcolor{gray!20}0.8776\\
& DctQim & 0.9749 & 0.9548 & 0.8623 & 0.8139 & 0.9121 & 0.8049 & 0.7073 & 0.9203 & 0.9278 & 0.5867 & 0.8465 \\
& \cellcolor{gray!20}TreeRing & \cellcolor{gray!20}0.9814 & \cellcolor{gray!20}0.9207 & \cellcolor{gray!20}0.8321 & \cellcolor{gray!20}0.7392 & \cellcolor{gray!20}0.9442 & \cellcolor{gray!20}0.8599 & \cellcolor{gray!20}0.8091 & \cellcolor{gray!20}0.9621 & \cellcolor{gray!20}0.9584 & \cellcolor{gray!20}0.6257 & \cellcolor{gray!20}0.8632 \\
& GaussianShading & 0.9882 & 0.9481 & 0.8745 & 0.7998 & 0.9217 & 0.8846 & 0.7850 & 0.9364 & 0.9512 & 0.6094 & 0.8699 \\
& \cellcolor{gray!20}W/O LAWMM & \cellcolor{gray!20}\underline{0.9980} & \cellcolor{gray!20}0.9623 & \cellcolor{gray!20}0.9352 & \cellcolor{gray!20}0.9147 & \cellcolor{gray!20}\underline{0.9795} & \cellcolor{gray!20}\underline{0.9592} & \cellcolor{gray!20}\textbf{0.9341} & \cellcolor{gray!20}\underline{0.9817} &  \cellcolor{gray!20}\underline{0.9706} & \cellcolor{gray!20}\underline{0.9895} & \cellcolor{gray!20}\underline{0.9625}\\
& Only CL & 0.9942 & 0.9537 & 0.9216 & 0.8864 & 0.9214 & 0.8745 & 0.8063 & 0.9426 & 0.9535  & 0.9635 & 0.9218\\
& \cellcolor{gray!20}Only VL & \cellcolor{gray!20}0.9828 & \cellcolor{gray!20}\underline{0.9765} & \cellcolor{gray!20}\underline{0.9433} & \cellcolor{gray!20}\underline{0.9172} & \cellcolor{gray!20}0.9670 & \cellcolor{gray!20}0.9148 & \cellcolor{gray!20}0.8592 & \cellcolor{gray!20}0.9521 &\cellcolor{gray!20}0.9676 & \cellcolor{gray!20}0.9787 & \cellcolor{gray!20}0.9459 \\
& KGMark & \textbf{0.9991} & \textbf{0.9810} & \textbf{0.9564} & \textbf{0.9207} & \textbf{0.9829} & \textbf{0.9669} & \underline{0.9320} &\textbf{0.9841} & \textbf{0.9809} & \textbf{0.9933}  & \textbf{0.9697}\\
\midrule
\multirow{8}{*}{MIND} 
& \cellcolor{gray!20}DwtDct & \cellcolor{gray!20}0.9793 & \cellcolor{gray!20}0.9563 & \cellcolor{gray!20}0.8827 & \cellcolor{gray!20}0.8161 & \cellcolor{gray!20}0.9334 & \cellcolor{gray!20}0.8413 & \cellcolor{gray!20}0.7610  &\cellcolor{gray!20}0.9358 &\cellcolor{gray!20}0.9291 & \cellcolor{gray!20}0.6348 & \cellcolor{gray!20}0.8669\\
& DctQim & 0.9785 & 0.9542 & 0.8703 & 0.8269 & 0.9046 & 0.8071& 0.6993  &0.9209 &0.9198 & 0.5708 & 0.8452\\
& \cellcolor{gray!20}TreeRing & \cellcolor{gray!20}0.9862 & \cellcolor{gray!20}0.9128 & \cellcolor{gray!20}0.8554 & \cellcolor{gray!20}0.8171 & \cellcolor{gray!20}0.9704 & \cellcolor{gray!20}0.8617 & \cellcolor{gray!20}0.7831 & \cellcolor{gray!20}0.9682 & \cellcolor{gray!20}0.9543 & \cellcolor{gray!20}0.5763 & \cellcolor{gray!20}0.8685 \\
& GaussianShading & 0.9903 & 0.9076 & 0.8455 & 0.7930 & 0.9334 & 0.8746 & 0.8284 & 0.9767 & 0.9681 & 0.5845 &0.8702  \\
& \cellcolor{gray!20}W/O LAWMM & \cellcolor{gray!20}\underline{0.9984} & \cellcolor{gray!20}\underline{0.9896} & \cellcolor{gray!20}\underline{0.9743} & \cellcolor{gray!20}\underline{0.9335} & \cellcolor{gray!20}\textbf{0.9895} & \cellcolor{gray!20}\underline{0.9609} & \cellcolor{gray!20}\underline{0.9328}  &\cellcolor{gray!20}\textbf{0.9915} &\cellcolor{gray!20}\underline{0.9864} & \cellcolor{gray!20}0.9705  &\cellcolor{gray!20}\underline{0.9727}\\
& Only CL & 0.9973 & 0.9792 & 0.9515 & 0.9287 & 0.9624 & 0.9074 & 0.8449  &0.9697 &0.9713 & \underline{0.9846}  & 0.9497\\
& \cellcolor{gray!20}Only VL & \cellcolor{gray!20}0.9956 & \cellcolor{gray!20}0.9814 & \cellcolor{gray!20}0.9658 & \cellcolor{gray!20}\textbf{0.9346} & \cellcolor{gray!20}0.9647 & \cellcolor{gray!20}0.9154 & \cellcolor{gray!20}0.8804  &\cellcolor{gray!20}0.9683 &\cellcolor{gray!20}0.9618 & \cellcolor{gray!20}\textbf{0.9861} & \cellcolor{gray!20}0.9554\\
& KGMark & \textbf{0.9987} & \textbf{0.9907} & \textbf{0.9751} & 0.9314 & \underline{0.9781} & \textbf{0.9726} & \textbf{0.9576}  &\underline{0.9849} &\textbf{0.9883} & 0.9842  &\textbf{0.9762} \\
\midrule
\multirow{8}{*}{Last-FM} 
& \cellcolor{gray!20}DwtDct & \cellcolor{gray!20}0.9801 & \cellcolor{gray!20}0.9636 & \cellcolor{gray!20}0.8834 & \cellcolor{gray!20}0.8229 & \cellcolor{gray!20}0.9570 & \cellcolor{gray!20}0.8353 & \cellcolor{gray!20}0.7415  & \cellcolor{gray!20}0.9596& \cellcolor{gray!20}0.9678& \cellcolor{gray!20}0.6407 & \cellcolor{gray!20}0.8752\\
& DctQim & 0.9842 & 0.9527 & 0.8773 & 0.8062 & 0.9083 & 0.7972 & 0.7125 &0.9144 &0.9161& 0.5938 & 0.8463\\
& \cellcolor{gray!20}TreeRing & \cellcolor{gray!20}0.9879 & \cellcolor{gray!20}0.9167 & \cellcolor{gray!20}0.8353 & \cellcolor{gray!20}0.7982 & \cellcolor{gray!20}0.9316 & \cellcolor{gray!20}0.9024 & \cellcolor{gray!20}0.8519 & \cellcolor{gray!20}0.9553 & \cellcolor{gray!20}0.9487 & \cellcolor{gray!20}0.6109 &\cellcolor{gray!20}0.8738  \\
& GaussianShading & 0.9795 & 0.9356 & 0.8658 & 0.8303 & 0.9519 & 0.9145 & 0.8667 & 0.9638 & 0.9594 & 0.6551 & 0.8922 \\
& \cellcolor{gray!20}W/O LAWMM & \cellcolor{gray!20}\textbf{0.9982} & \cellcolor{gray!20}\underline{0.9929} & \cellcolor{gray!20}\textbf{0.9857} & \cellcolor{gray!20}\textbf{0.9468} & \cellcolor{gray!20}0.9785 & \cellcolor{gray!20}\textbf{0.9420} & \cellcolor{gray!20}\textbf{0.9057}  &\cellcolor{gray!20}\underline{0.9794} &\cellcolor{gray!20}\textbf{0.9852} & \cellcolor{gray!20}\underline{0.9970}  & \cellcolor{gray!20}\textbf{0.9711}\\
& Only CL & 0.9962 & 0.9893 & 0.9367 & 0.9073 & 0.9773 & 0.9206 & 0.8737  &0.9156 &0.9205 & 0.9901 & 0.9427\\
& \cellcolor{gray!20}Only VL & \cellcolor{gray!20}0.9929 & \cellcolor{gray!20}0.9726 & \cellcolor{gray!20}0.9308 & \cellcolor{gray!20}0.9034 & \cellcolor{gray!20}\textbf{0.9824} & \cellcolor{gray!20}0.9275 & \cellcolor{gray!20}0.8661  &\cellcolor{gray!20}0.9053 &\cellcolor{gray!20}0.9112 & \cellcolor{gray!20}0.9912 &\cellcolor{gray!20}0.9383 \\
& KGMark & \underline{0.9976} & \textbf{0.9973} & \underline{0.9844} & \underline{0.9421} & \underline{0.9796} & \underline{0.9350} & \underline{0.9031} &\textbf{0.9886} & \underline{0.9814} & \textbf{0.9977}  & \underline{0.9707}\\
\bottomrule
\end{tabular}%
}
\vspace{-0.4em}
\label{tab: detect robust}
\end{table*}

\vspace{-0.4em}
\section{Experiments}
The experimental evaluation of \sysname assesses its effectiveness in terms of watermark \textbf{transparency}, \textbf{detectability}, and \textbf{robustness} across various attack scenarios and datasets.

\vspace{-0.4em}
\subsection{Experiment setup}
\vspace{-0.4em}

\textbf{Datasets.}
We evaluate our approach using three public datasets representing diverse real-world scenarios: Last-FM (music)~\cite{ccano2017music}, MIND (news)~\cite{wu-etal-2020-mind}, and Alibaba-iFashion (e-commerce)~\cite{10.1145/3292500.3330652}. Table~\ref{tab: ds} provides a summary of these datasets.

\begin{table}[ht]
\centering
\vspace{-1em}
\caption{We use three KG datasets from different domains.}
\resizebox{\columnwidth}{!}{%
\begin{tabular}{cccc}
\toprule
\textbf{Name} & \textbf{Entities} & \textbf{Relations} & \textbf{Triples}  \\
\midrule
Alibaba-iFashion (AliF) & 59,156 & 51 & 279,155  \\
MIND & 24,733 & 512 & 148,568 \\
Last-FM	 & 58,266 & 9 & 464,567	 \\
\bottomrule
\end{tabular}
}
\label{tab: ds}
\vspace{-1.5em}
\end{table}

\textbf{Variants.}
Since \sysname is the first watermarking scheme for Knowledge Graph Embedding (KGE), we explored multiple variants in the experiments.
\ding{182} \textbf{W/O LAWMM}: Using a fixed watermark mask matrix with the same density, instead of applying the Learnable Adaptive Watermark Mask Matrix (LAWMM).
\ding{183} \textbf{Only CL}: Applying the watermark exclusively in the Community Layer, selecting specific vertices or vertex groups within each community.
\ding{184} \textbf{Only VL}: Applying the watermark exclusively in the Vertex Layer, selecting multiple vertices or vertex groups within a specific community.
\ding{185} \textbf{W/O Watermark (Control)}: Reconstructing the knowledge graph without any watermark.

\textbf{Baselines.}
As the first watermarking framework designed for diffusion KGE, we introduce four additional baselines to validate \sysname's effectiveness.
\textbf{TreeRing}~\cite{NEURIPS2023_b54d1757} and \textbf{GaussianShading}~\cite{Yang_2024_CVPR} are node-level watermarking methods designed for diffusion models (Images), embedding watermarks by replacing 5\% of nodes in the graph. \textbf{DwtDct}~\cite{cox2007digital} and \textbf{DctQim}~\cite{chen2001class} are classical watermarking techniques that modify transformed coefficients to balance imperceptibility and robustness.

\textbf{Implementation details.} 
We first employ the RotatE~\cite{sun2019rotate} model to embed the knowledge graph, with an embedding dimension of 4096. 
Our watermarking method is applied to the above-processed datasets, and a subsequent series of related experiments is carried out. All experiments are conducted on a single NVIDIA A800.

\vspace{-0.5em}
\subsection{Detectability}
\vspace{-0.4em}

To evaluate the detectability of \sysname, we calculate the false positive rate (FPR), true positive rate (TPR), and their corresponding average AUC values, as summarized in Table~\ref{tab: detect robust}. 
The evaluations are conducted under a configuration where the DDIM inference steps are set to 75, and the predefined significance level is fixed at $5 \times 10^{-5}$. 
To further examine the impact of significance levels and DDIM inference steps on watermark detectability, we perform controlled experiments by varying these parameters.

\begin{table*}[t]
\centering
\scriptsize
\caption{Watermark Transparency. We evaluate the transparency of watermarking from two dimensions: the \textbf{similarity} of knowledge graph embedding before and after watermarking and the \textbf{quality} of watermarked knowledge graph. We \underline{\textbf{mark}} the best transparency score only for the four baselines and two \sysname variants, as the original KG and W/O Watermark serve as control groups.}
\resizebox{\textwidth}{!}{%
\begin{tabular}{l|l|ccc|cccc}
\toprule
\multirow{2}{*}{\textbf{Datasets}} & \multirow{2}{*}{\textbf{Method}} & \multicolumn{3}{c|}{\textbf{Cosine Similarity $\uparrow$}} & \multicolumn{4}{c}{\textbf{KG Quality Metric @ 75 Steps}} \\
\cmidrule(lr){3-5} \cmidrule(lr){6-9}
& & \cellcolor{gray!10}50 Steps & \cellcolor{gray!20}65 Steps & \cellcolor{gray!30}75 Steps & GMR $\downarrow$ & HMR $\downarrow$ & AMR $\downarrow$ & Hits@10 $\uparrow$ \\
\midrule
\multirow{8}{*}{AliF} 
 & Original KG& - & - & - & 1.828 & 1.162 & 135.459 & 0.8980 \\
 & \cellcolor{gray!20}W/O Watermark & \cellcolor{gray!20}0.7971 & \cellcolor{gray!20}0.8797 & \cellcolor{gray!20}0.9674 & \cellcolor{gray!20}3.026 & \cellcolor{gray!20}1.579 & \cellcolor{gray!20}141.412 & \cellcolor{gray!20}0.8318 \\
 & DwtDct & 0.7215 & 0.7928 & 0.8251 & 5.096 & 1.699 & 157.036 & 0.6933 \\
 & \cellcolor{gray!20}DctQim & \cellcolor{gray!20}0.7509 & \cellcolor{gray!20}0.7633 & \cellcolor{gray!20}0.7653 & \cellcolor{gray!20}5.104 & \cellcolor{gray!20}1.654 & \cellcolor{gray!20}161.142 & \cellcolor{gray!20}0.7385 \\
 & TreeRing & \underline{0.7761} & \textbf{0.8431} & \underline{0.9071} & 3.928& \underline{1.618} & 152.634 & \underline{0.8017} \\
 & \cellcolor{gray!20}GaussianShading & \cellcolor{gray!20}0.2879 & \cellcolor{gray!20}0.3226 & \cellcolor{gray!20}0.3538 & \cellcolor{gray!20}6.641 & \cellcolor{gray!20}1.798 & \cellcolor{gray!20}172.813 & \cellcolor{gray!20}0.5137 \\
 & W/O LAWMM & 0.7662 & 0.7838 & 0.8643 & \underline{3.457} & 1.624 & \underline{147.305} & 0.7871\\
 & \cellcolor{gray!20}KGMark & \cellcolor{gray!20}\textbf{0.7839} & \cellcolor{gray!20}\underline{0.8309} & \cellcolor{gray!20}\textbf{0.9482} & \cellcolor{gray!20}\textbf{3.046} & \cellcolor{gray!20}\textbf{1.580} & \cellcolor{gray!20}\textbf{141.904} & \cellcolor{gray!20}\textbf{0.8296} \\
\midrule
\multirow{8}{*}{MIND} & Original KG& - & - & - & 7.197 & 1.975 & 155.656 & 0.6649 \\
 & \cellcolor{gray!20}W/O Watermark & \cellcolor{gray!20}0.7345 & \cellcolor{gray!20}0.8290 & \cellcolor{gray!20}0.9579 & \cellcolor{gray!20}9.853 & \cellcolor{gray!20}2.215 & \cellcolor{gray!20}168.906 & \cellcolor{gray!20}0.5734 \\
 & DwtDct & 0.7831 & \underline{0.8244} & 0.8312 & 11.328 & 2.328 & 188.992 & 0.5216 \\
 & \cellcolor{gray!20}DctQim & \cellcolor{gray!20}0.7549 & \cellcolor{gray!20}0.7574 & \cellcolor{gray!20}0.7703 & \cellcolor{gray!20}12.037 & \cellcolor{gray!20}2.753 & \cellcolor{gray!20}205.483 & \cellcolor{gray!20}0.4835\\
 & TreeRing & \underline{0.7976} & 0.8108 & 0.8581 & 11.102 & 2.297& 182.925 & 0.5108 \\
 & \cellcolor{gray!20}GaussianShading & \cellcolor{gray!20}0.2843 & \cellcolor{gray!20}0.3196 &  \cellcolor{gray!20}0.3728 & \cellcolor{gray!20}14.523 & \cellcolor{gray!20}3.312 & \cellcolor{gray!20}234.064 & \cellcolor{gray!20}0.3940 \\
 & W/O LAWMM& 0.7469& 0.8235 & \underline{0.8902} & \underline{10.819} & \underline{2.238} & \underline{173.194} & \underline{0.5332} \\
 & \cellcolor{gray!20}KGMark & \cellcolor{gray!20}\textbf{0.8083} & \cellcolor{gray!20}\textbf{0.8533} & \cellcolor{gray!20}\textbf{0.9397} & \cellcolor{gray!20}\textbf{10.508} & \cellcolor{gray!20}\textbf{2.226} & \cellcolor{gray!20}\textbf{169.305} & \cellcolor{gray!20}\textbf{0.5683} \\
\midrule
\multirow{8}{*}{Last-FM} & Original KG & -& -& -& 3.571 & 1.202 & 1711.695 & 0.8436 \\
 & \cellcolor{gray!20}W/O Watermark & \cellcolor{gray!20}0.8293 & \cellcolor{gray!20}0.9157 & \cellcolor{gray!20}0.9410 & \cellcolor{gray!20}4.455 & \cellcolor{gray!20}1.452 & \cellcolor{gray!20}1715.907 & \cellcolor{gray!20}0.8431 \\
 & DwtDct & 0.7215 & 0.7928 & 0.8433 & 4.519 & 1.502 & 1734.823 & 0.8221 \\
 & \cellcolor{gray!20}DctQim & \cellcolor{gray!20}0.7509 & \cellcolor{gray!20}0.7633 & \cellcolor{gray!20}0.7679 & \cellcolor{gray!20}5.139 & \cellcolor{gray!20}1.704 & \cellcolor{gray!20}2043.249 & \cellcolor{gray!20}0.7264\\
 & TreeRing & 0.7262 & 0.7896 & 0.8364  & 4.733 & 1.652 & 1772.961 & 0.8149 \\
 & \cellcolor{gray!20}GaussianShading &\cellcolor{gray!20}0.3252 &\cellcolor{gray!20}0.3649 &\cellcolor{gray!20}0.4184 &\cellcolor{gray!20}6.349 &\cellcolor{gray!20}2.016 &\cellcolor{gray!20}2192.492 &\cellcolor{gray!20}0.6463 \\
 & W/O LAWMM& \underline{0.7628}& \underline{0.8809} & \underline{0.8948} & \underline{4.456} & \underline{1.452} & \underline{1716.828} & \underline{0.8430} \\
 & \cellcolor{gray!20}KGMark & \cellcolor{gray!20}\textbf{0.8876} & \cellcolor{gray!20}\textbf{0.9051} & \cellcolor{gray!20}\textbf{0.9161} & \cellcolor{gray!20}\textbf{4.455} & \cellcolor{gray!20}\textbf{1.452} & \cellcolor{gray!20}\textbf{1716.365} & \cellcolor{gray!20}\textbf{0.8430} \\
\midrule
\end{tabular}%
}
\label{tab: transparency}
\vspace{-0.6em}
\end{table*}

\begin{figure*}[ht]
    \centering
    \begin{subfigure}[t]{0.30\textwidth}
        \centering
        \includegraphics[width=\textwidth]{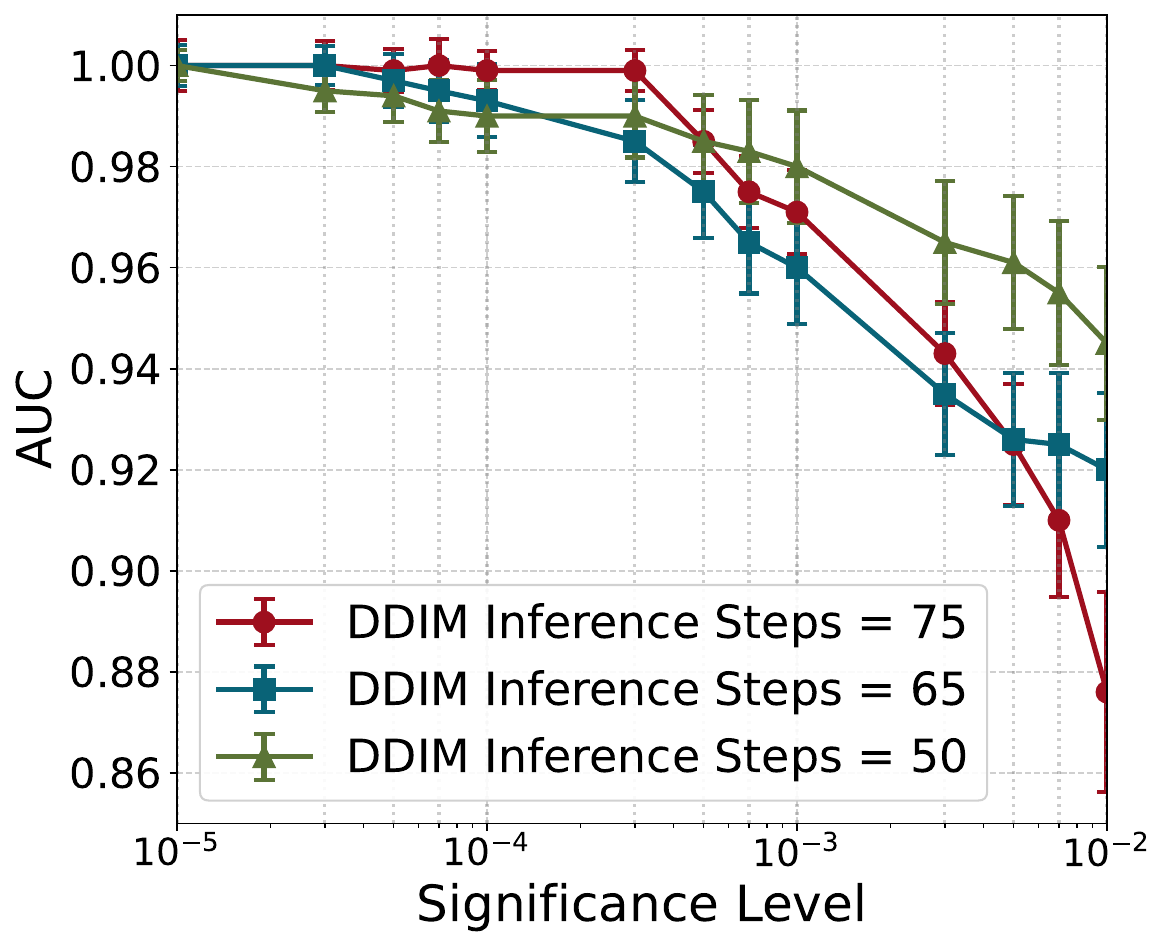}
        \caption{Watermark Detectability}
        \label{fig:detect}
    \end{subfigure}
    \hfill
    \begin{subfigure}[t]{0.33\textwidth}
        \centering
        \includegraphics[width=\textwidth]{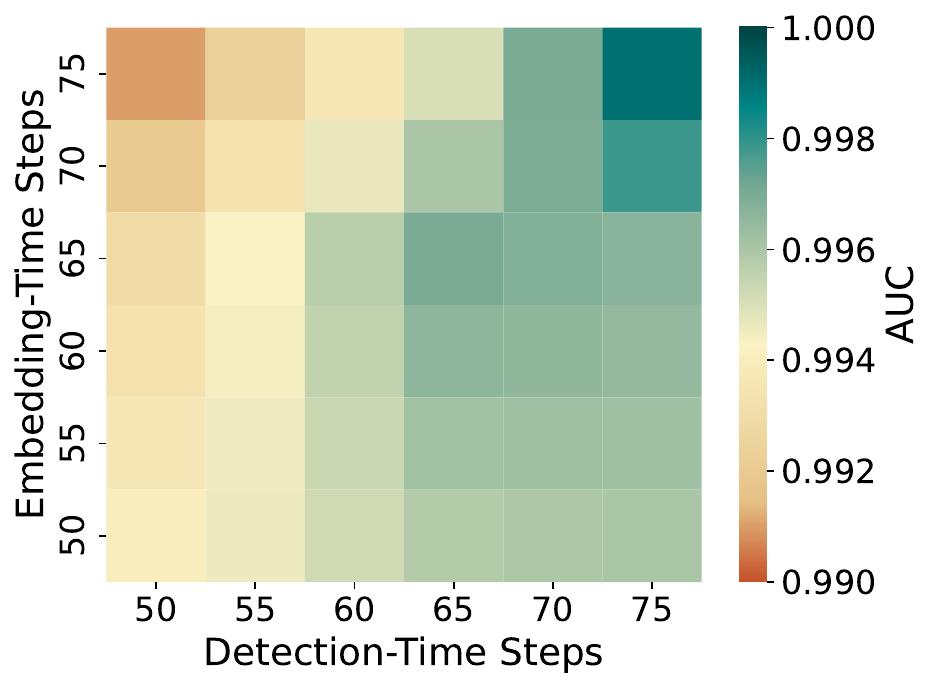}
        \caption{Ablation on DDIM Steps}
        \label{fig:hotmap}
    \end{subfigure}
    \hfill
    \begin{subfigure}[t]{0.33\textwidth}
        \centering
        \includegraphics[width=\textwidth]{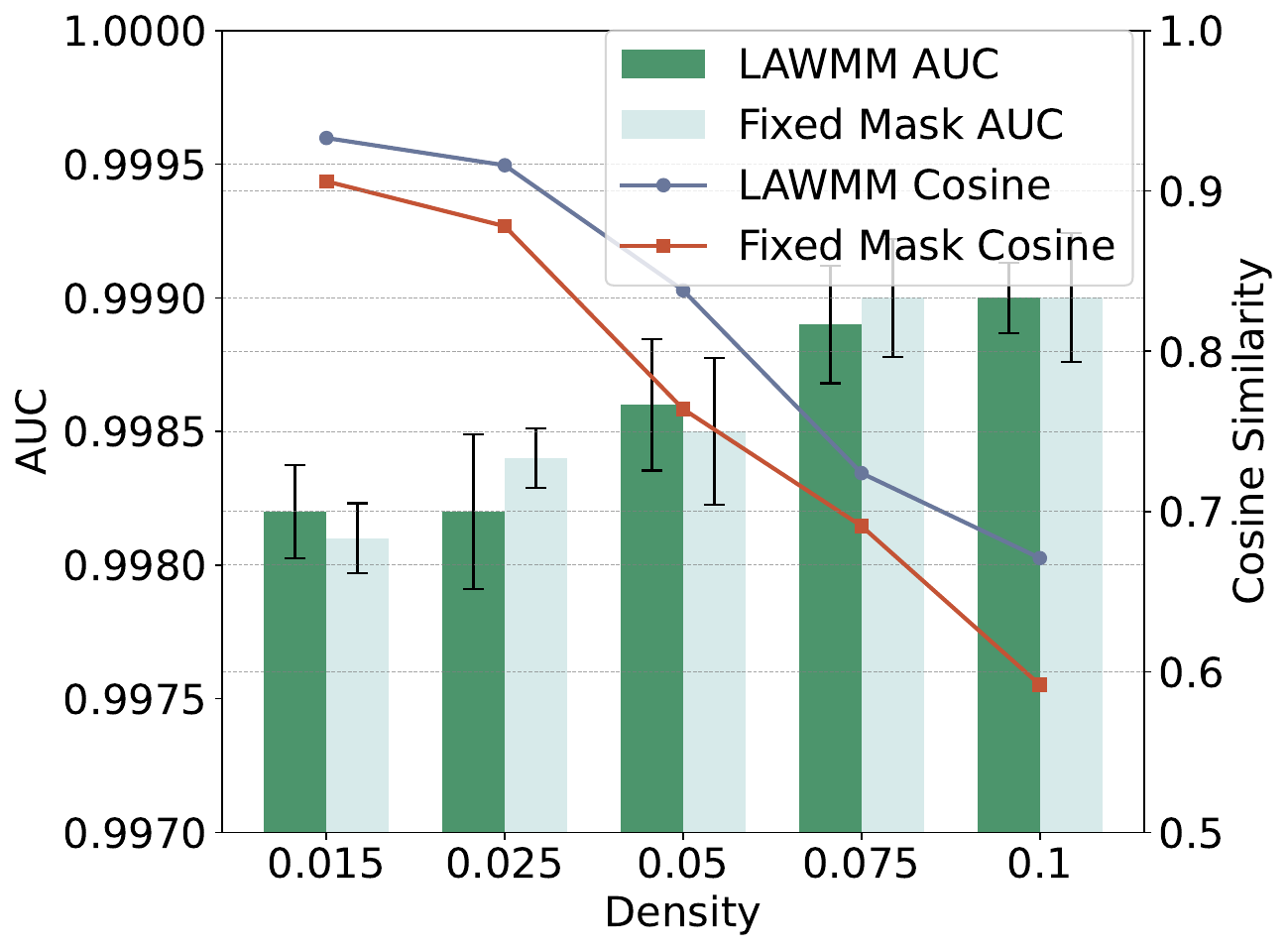}
        \caption{Effect of Mask Density}
        \label{fig:density}
    \end{subfigure}
    \vspace{-0.2em}
    \caption{Ablation. (a) AUC under varying significance levels and DDIM inference steps. (b) Different DDIM inference steps during embedding and detection. (c) Effect of watermark mask density on all datasets, with DDIM steps set to 75 and significance level $5e{-}5$.}
    \label{fig:case}
    \vspace{-1em}
\end{figure*}


Table~\ref{tab: detect robust} summarizes the detection performance of \sysname under various attack scenarios. 
Across all settings, \sysname consistently maintains high detectability.
We also evaluate an ablated variant (W/O LAWMM) that removes the Learnable Adaptive Watermark Mask, which is designed to preserve transparency while balancing embedding detectability. Interestingly, this variant exhibits slightly higher detectability in some cases, likely due to stronger watermark signal strength without the masking constraint. 
However, the full \sysname still achieves comparable or superior performance overall.
This result also highlights the effectiveness of LAWMM in achieving a favorable trade-off between watermark detectability and semantic transparency.

\textbf{Obs.\ding{182} \sysname's detectability remains consistently high with optimal DDIM steps and significance levels.}
Figure~\ref{fig:detect} illustrates the impact of significance levels and DDIM inference steps on the AUC values. 
Within the significance level range of $1e-5 \sim 1e-4$, increasing the DDIM steps significantly enhances detection performance by reducing watermark information loss. 
Meanwhile, the AUC values remain exceptionally stable, approaching 1 for configurations with higher steps, highlighting \sysname's robust and consistent performance under suitable conditions.

However, for higher significance levels $1e-3 \sim 1e-2$, the AUC values drop more sharply. 
This decline is attributed to higher inference steps reducing p-values for negative samples, thereby increasing the false positive rate (FPR) and accelerating the drop in detection accuracy.

\textbf{Obs.\ding{183} Increasing detection-stage DDIM steps boosts detectability, while stage alignment further enhances performance.}
Results in Figure~\ref{fig:hotmap} demonstrate that higher DDIM steps in the detection stage yield marginal improvements in AUC, with values consistently exceeding 0.996 across most configurations. 
For example, as detection steps increase from lower to higher ranges, the AUC improves slightly, showcasing the positive impact of precise parameter tuning during detection.

However, inconsistencies between embedding and detection steps introduce minor declines in performance, as seen in higher-step configurations where AUC decreases from $0.99785 \sim 0.9967$. 
This highlights the critical importance of alignment between the two stages. 
Notably, the detection stage exerts a more significant influence on watermark retrieval accuracy, reinforcing the need for fine-grained control at this stage to achieve optimal detectability.

\vspace{-0.4em}
\subsection{Transparency}
\vspace{-0.4em}

We evaluate the transparency of \sysname, which measures the extent to which the watermark preserves the structural and functional integrity of the original KGE.

\textbf{Obs.\ding{184} \sysname achieves strong transparency by minimally impacting KG structure and usability.}
As shown in Table~\ref{tab: transparency}, KGMark achieves high cosine similarity scores (e.g., 0.9482 for AliF at 75 steps) compared to the non-watermarked version, indicating that the watermark does not significantly distort the KG's inherent relationships. 
This demonstrates that the watermark is seamlessly integrated without disrupting the KG's core structure.

To further evaluate the KG's quality post-watermarking, we measure forecast task performance using GMR, HMR, AMR, and Hits@10 metrics. KGMark's results are nearly on par with the original KG and the non-watermarked version. For example, in AliF, KGMark achieves a Hits@10 score of 0.8296, slightly lower than the non-watermarked score of 0.8318. Similar trends are observed across MIND and Last-FM, where KGMark consistently outperforms the version without LAWMM. These findings confirm that KGMark preserves the KG's utility for downstream tasks, ensuring the watermarked KG remains functional and reliable.

Experimental results in Figure~\ref{fig:density} show that the density of the watermark mask matrix also influences watermark transparency. 
Higher density causes a slight increase in detectability. However, this comes at the cost of transparency, as cosine similarity (LAWMM\_cosine) drops from 0.933 to 0.671, reflecting degradation in KG integrity. 
This trade-off underscores the need to balance density for optimal detectability and transparency.

\vspace{-0.4em}
\subsection{Robustness}
\vspace{-0.4em}
We next assess the robustness of \sysname, focusing on its ability to withstand a wide range of post-editing attacks without compromising watermark integrity.
To evaluate the robustness of our watermarking scheme, we systematically analyze its performance under five post-editing attacks with varying intensities: Gaussian noise injection, Gaussian smoothing\cite{hu2024transfer}, relation alteration, triple deletion, and graph isomorphism variation. 
Attack intensity is quantified by the proportion of affected entities or triples, ranging from 10\% to 50\%, simulating real-world adversarial scenarios.
We further incorporate two stronger adversarial attacks into our evaluation: NEA~\cite{bojchevski2019adversarial}, a graph poisoning attack targeting node embeddings, and the L2 Metric attack~\cite{bhardwaj2021adversarial}, which perturbs embeddings via instance attribution analysis. 
These attacks introduce fine-grained and targeted perturbations, enabling rigorous robustness evaluation under realistic and challenging adversarial settings.

\begin{table}[ht]
\centering
\caption{Watermark Detectability \& Robustness (Gaussian Noise \& Smoothing). We impose two common attacks against images on KGE and evaluate the robustness of these two attacks under different attack intensities.}
\resizebox{\columnwidth}{!}{%
\begin{tabular}{l|l|cccccc}
\toprule
\multirow{2}{*}{\textbf{Datasets}} & \multirow{2}{*}{\textbf{Method}} & \multicolumn{3}{c}{\textbf{Gaussian Noise}} & \multicolumn{3}{c}{\textbf{Smoothing}} \\
\cmidrule(lr){3-5} \cmidrule(lr){6-8}
& & \cellcolor{gray!10}10\% & \cellcolor{gray!20}30\% & \cellcolor{gray!30}50\% & \cellcolor{gray!10}10\% & \cellcolor{gray!20}30\% & \cellcolor{gray!30}50\% \\
\midrule
\multirow{8}{*}{AliF} 
& \cellcolor{gray!20}DwtDct & \cellcolor{gray!20}0.98 & \cellcolor{gray!20}0.89 & \cellcolor{gray!20}0.86 & \cellcolor{gray!20}0.94 & \cellcolor{gray!20}0.90 & \cellcolor{gray!20}0.81 \\
& DwtQim & 0.93 & 0.84 & 0.69 & 0.90 & 0.82 & 0.77 \\
& \cellcolor{gray!20}TR & \cellcolor{gray!20}0.92 & \cellcolor{gray!20}0.86 & \cellcolor{gray!20}0.81 & \cellcolor{gray!20}0.92 & \cellcolor{gray!20}0.86 & \cellcolor{gray!20}0.78 \\
& GaussianShading & 0.94 & 0.91 & 0.89 & 0.90 & 0.83 & 0.79  \\
& \cellcolor{gray!20}W/O LAWMM & \cellcolor{gray!20}\underline{0.98} & \cellcolor{gray!20}\underline{0.95} & \cellcolor{gray!20}\textbf{0.92} & \cellcolor{gray!20}\textbf{0.95} & \cellcolor{gray!20}\underline{0.91} & \cellcolor{gray!20}\underline{0.88} \\
& Only CL & 0.98 & 0.94 & 0.90 & 0.93 & 0.90 & 0.84 \\
& \cellcolor{gray!20}Only VL & \cellcolor{gray!20}0.96 & \cellcolor{gray!20}0.92 & \cellcolor{gray!20}0.87 & \cellcolor{gray!20}0.91 & \cellcolor{gray!20}0.87 & \cellcolor{gray!20}0.82\\
& KGMark & \textbf{0.98} & \textbf{0.96} & \underline{0.91} & \underline{0.94} & \textbf{0.92} & \textbf{0.89} \\
\midrule
\multirow{8}{*}{MIND} 
& \cellcolor{gray!20}DwtDct & \cellcolor{gray!20}0.96 & \cellcolor{gray!20}0.90 & \cellcolor{gray!20}0.83 & \cellcolor{gray!20}\underline{0.96} & \cellcolor{gray!20}0.87 & \cellcolor{gray!20}0.81 \\
& DwtQim & 0.93 & 0.81 & 0.72 & 0.94 & 0.85 & 0.79 \\
& \cellcolor{gray!20}TR & \cellcolor{gray!20}0.91 & \cellcolor{gray!20}0.86 & \cellcolor{gray!20}0.77 &\cellcolor{gray!20} 0.94 & \cellcolor{gray!20}0.91 &\cellcolor{gray!20}0.83\\
& GaussianShading & 0.94 & 0.89 & 0.85 & 0.90 & 0.86 & 0.83 \\
& \cellcolor{gray!20}W/O LAWMM & \cellcolor{gray!20}\underline{0.98} & \cellcolor{gray!20}\underline{0.95} & \cellcolor{gray!20}\underline{0.91} & \cellcolor{gray!20}0.95 & \cellcolor{gray!20}\underline{0.92} & \cellcolor{gray!20}\underline{0.89} \\
& Only CL & 0.98 & 0.93 & 0.90 & 0.96 & 0.92 & 0.87 \\
& \cellcolor{gray!20}Only VL & \cellcolor{gray!20}0.94 & \cellcolor{gray!20}0.93 & \cellcolor{gray!20}0.91 & \cellcolor{gray!20}0.95 & \cellcolor{gray!20}0.91 & \cellcolor{gray!20}0.89 \\
& KGMark & \textbf{0.99} & \textbf{0.96} & \textbf{0.92} & \textbf{0.96} & \textbf{0.93} & \textbf{0.90} \\
\midrule
\multirow{8}{*}{Last-FM} 
& \cellcolor{gray!20}DwtDct & \cellcolor{gray!20}0.96 & \cellcolor{gray!20}0.90 & \cellcolor{gray!20}0.85 & \cellcolor{gray!20}0.95 & \cellcolor{gray!20}0.85 &\cellcolor{gray!20}0.77 \\
& DwtQim & 0.95 & 0.84 & 0.74 & 0.95 & 0.92 & 0.82 \\
& \cellcolor{gray!20}TR & \cellcolor{gray!20}0.93 &\cellcolor{gray!20}0.89 &\cellcolor{gray!20}0.86 & \cellcolor{gray!20}0.92 & \cellcolor{gray!20}0.89 & \cellcolor{gray!20}0.85 \\
& GaussianShading & 0.96 & 0.92 & 0.89 & 0.92 & 0.89 & 0.86  \\
& \cellcolor{gray!20}W/O LAWMM & \cellcolor{gray!20}\underline{0.99} & \cellcolor{gray!20}\underline{0.96} & \cellcolor{gray!20}\underline{0.93} & \cellcolor{gray!20}\underline{0.95} & \cellcolor{gray!20}\underline{0.93} & \cellcolor{gray!20}\underline{0.91} \\
& Only CL & 0.97 & 0.96 & 0.92 & 0.93 & 0.91 & 0.89 \\
& \cellcolor{gray!20}Only VL & \cellcolor{gray!20}0.98 & \cellcolor{gray!20}0.96 & \cellcolor{gray!20}0.93 & \cellcolor{gray!20}0.94 & \cellcolor{gray!20}0.90 & \cellcolor{gray!20}0.86\\
& KGMark & \textbf{0.99} & \textbf{0.97} & \textbf{0.93} & \textbf{0.95} & \textbf{0.93} & \textbf{0.91} \\
\bottomrule
\end{tabular}%
}
\label{tab: robust_gaussian_smoothing}
\vspace{-0.6em}
\end{table}

\textbf{Obs.\ding{185} \sysname's hierarchical embedding ensures robustness by mitigating both local perturbations and structural disruptions.}
As shown in Table~\ref{tab: robust_gaussian_smoothing} and Table~\ref{tab: detect robust}, the robustness of the watermark varies significantly across method variants. 
The standalone Community Layer (Only CL) and Vertex Layer (Only VL) exhibit weaker resilience compared to the full \sysname. 
For instance, under triple deletion attacks on AliF at 50\% intensity, Only CL and Only VL achieve AUC scores of 0.8063 and 0.8592, respectively, while \sysname maintains a robust score of 0.9320. 
This underscores the necessity of combining coarse-grained (Community Layer) and fine-grained (Vertex Layer) embedding strategies. 
Notably, the variant without LAWMM (W/O LAWMM) demonstrates comparable robustness to \sysname, as LAWMM is primarily designed for transparency and does not compromise robustness. 

Our analysis reveals that \sysname is particularly sensitive to triple deletion and high-intensity smoothing attacks while maintaining strong robustness against other attack types. 
Under triple deletion attacks on AliF, the AUC drops from 0.9829 at 10\% intensity to 0.9320 at 50\% intensity, reflecting a significant but manageable decline. 
Similarly, under smoothing attacks, the AUC decreases from 0.94 at 10\% intensity to 0.89 at 50\% intensity on AliF. 
In contrast, \sysname demonstrates exceptional resilience to Gaussian noise and relation alteration, achieving an AUC of 0.93 at 50\% intensity under Gaussian noise attacks on Last-FM. 
These results validate that \sysname effectively balances robustness and usability, even under aggressive post-editing attacks.

\vspace{-0.4em}
\subsection{Case Study}
\vspace{-0.4em}
We conduct a case study on the downstream task of news recommendation using the MIND dataset, comparing the performance of knowledge graph embeddings with and without the proposed watermark. 
The experiments are carried out using the Deep Knowledge-Aware Network (DKN) model~\cite{wang2018dkn}, which has been trained for 10 epochs on the MIND dataset. 
To simulate a controlled user profile, we restrict the user's click history by focusing on a strong interest in sports news. Additionally, we randomly sample 5 news items from the model's output for evaluation.
\vspace{-0.5em}
\begin{figure}[ht]
    \centering
    \includegraphics[width=0.48\textwidth]{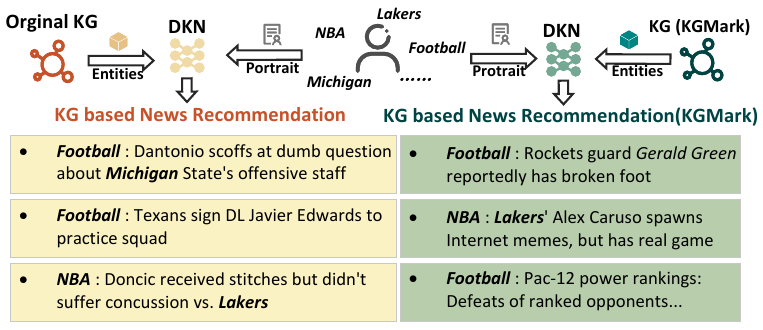}
    \vspace{-2em}
    \caption{Case Study. Workflow and representative results.}
    \label{fig:casestudy}
    \vspace{-1.3em}
\end{figure}

The results, shown in Figure~\ref{fig:casestudy}, highlight that the content and subcategories of targeted recommendations remain consistent with and without watermarking. Despite the watermark, the recommendation system focuses on sports news and identifies trending topics, such as the "Pac-12 power rankings." Entity recognition remains stable, with consistent identification of sports figures, teams, and locations across both versions. Additionally, the recommendation logic and hot topic recommendations remain unchanged, indicating no impact from the watermarking process.

\vspace{-0.4em}
\section{Discussion}
\vspace{-0.4em}
\subsection{Variational Autoencoder}
\vspace{-0.4em}


\sysname's implementation employed a VAE based on the Relational Graph Attention Network (RGAT)~\cite{busbridge2019relationalgraphattentionnetworks}, which extends GAT~\cite{veličković2018graphattentionnetworks} by incorporating relational dependencies. Given a KG $\mathcal{G} = (\mathcal{V}, \mathcal{E}, \mathcal{R})$, RGAT updates node representations as:  
\vspace{-0.6em}
\begin{equation}
\mathbf{h}_v^{(l+1)} = \sigma \left( \sum_{r \in \mathcal{R}} \sum_{u \in \mathcal{N}r(v)} \alpha_{vu}^r \mathbf{W}^r \mathbf{h}_u^{(l)} \right),
\label{eq:rgat_update}
\end{equation}
where $\mathcal{N}_r(v)$ denotes neighbors connected via relation $r$, $\mathbf{W}^r$ is a relation-specific transformation, and $\alpha_{vu}^r$ is the learned attention weight.  

Relational graph neural networks~\cite{schlichtkrull2017modelingrelationaldatagraph} (e.g., RGAT, RGCN) better enhance the expressiveness of the latent space $\mathbf{z} \sim q_{\phi}(\mathbf{z} | \mathcal{G})$by capturing structural dependencies. However, our experiments show that relation-based models are vulnerable to adversarial attacks, which disrupt attention weights and cause representation shifts:  
\vspace{-0.4em}
\begin{equation}  
\|\tilde{\mathbf{h}}_v - \mathbf{h}_v\|_2^2 \approx \sum_{r \in \mathcal{R}} \left( \sum_{u \in \mathcal{N}_r(v) \setminus \tilde{\mathcal{N}}_r(v)} (\alpha_{vu}^r \mathbf{W}^r \mathbf{h}_u)^2 \right)
\label{eq:perturbation}  
\end{equation}  
where $\tilde{\mathcal{N}}_r(v)$ is the modified neighborhood. This degradation also impacts VAE reconstruction, reducing watermark detectability.  
Using non-relational models such as GCN or GAT mitigates this vulnerability but lowers latent space expressiveness, leading to suboptimal reconstruction:  
\begin{equation}  
D_{\mathrm{KL}}(q_{\phi}(\mathbf{z} | \mathcal{G}) \| p(\mathbf{z})) \uparrow, \quad \mathbb{E}_{q_{\phi}(\mathbf{z} | \mathcal{G})}[\log p_{\theta}(\mathcal{G} | \mathbf{z})] \downarrow.  
\label{eq:vae_loss}  
\end{equation}  
Thus, relational modeling improves representation quality but increases attack susceptibility, requiring a balance between robustness and expressiveness.
\vspace{-0.4em}
\subsection{Embedding Dimensions and DDIM Generality}
\vspace{-0.4em}
The dimensionality of embeddings affects watermark detectability and robustness, requiring a balance with task performance. While DDIM inversion is central to our method, compatibility with advanced samplers remains an open problem. Future work will explore conditional sampling to improve reconstruction quality and watermark transparency.

\vspace{-0.6em}
\section{Conclusion}
\vspace{-0.4em}
In this paper, we introduce \sysname, a novel watermarking scheme for knowledge graphs, leveraging diffusion models to ensure robustness, transparency, and detectability. 
Our method addresses key challenges in graph integrity, offering a secure solution that maintains semantic fidelity even in dynamic environments.
\sysname provides a foundation for securing the integrity and ownership of synthetic knowledge graphs, with potential applications spanning academic research to commercial deployments in fields like recommendation systems and GraphRAG.

\section*{Acknowledgements}
We would like to thank the anonymous reviewers for their valuable feedback. We are also grateful to the open-source community for providing benchmark datasets and foundational libraries that made this research possible. This work was supported in part by the National Natural Science Foundation of China~(No.62471064), the National Natural Science Foundation of China~(No.62301072) and the National Natural Science Foundation of China~(No.62271066).

\section*{Impact Statement}
This work proposes KGMark, the first diffusion watermarking framework for knowledge graphs, aiming to enhance intellectual property protection and traceability in AI-generated KGE data.  
Given the increasing reliance on synthetic KGs in downstream applications such as recommendation systems, semantic search, and language model augmentation, our method enables robust, transparent, and detectable watermarking at the embedding level.  
We do not foresee immediate negative societal impact from this work. However, we acknowledge that any technology for ownership control may raise ethical concerns if misused, such as improper surveillance or censorship. To mitigate this, we focus strictly on benign applications such as integrity verification and content provenance.


\bibliography{references}

\begin{thebibliography}{69}
\providecommand{\natexlab}[1]{#1}
\providecommand{\url}[1]{\texttt{#1}}
\expandafter\ifx\csname urlstyle\endcsname\relax
  \providecommand{\doi}[1]{doi: #1}\else
  \providecommand{\doi}{doi: \begingroup \urlstyle{rm}\Url}\fi

\bibitem[An et~al.(2025)An, Ding, Rabbani, Agrawal, Xu, Deng, Zhu, Mohamed, Wen, Goldstein, and Huang]{10.5555/3692070.3692129}
An, B., Ding, M., Rabbani, T., Agrawal, A., Xu, Y., Deng, C., Zhu, S., Mohamed, A., Wen, Y., Goldstein, T., and Huang, F.
\newblock Waves: benchmarking the robustness of image watermarks.
\newblock In \emph{Proceedings of the 41st International Conference on Machine Learning}. JMLR.org, 2025.

\bibitem[Asnani et~al.(2024)Asnani, Collomosse, Bui, Liu, and Agarwal]{asnani2024promark}
Asnani, V., Collomosse, J., Bui, T., Liu, X., and Agarwal, S.
\newblock Promark: Proactive diffusion watermarking for causal attribution.
\newblock In \emph{Proceedings of the IEEE/CVF Conference on Computer Vision and Pattern Recognition}, pp.\  10802--10811, 2024.

\bibitem[Barman et~al.(2024)Barman, Sharma, Aziz, Bajpai, Biswas, Sharma, Jain, Chadha, Sheth, and Das]{barman2024brittleness}
Barman, N.~R., Sharma, K., Aziz, A., Bajpai, S., Biswas, S., Sharma, V., Jain, V., Chadha, A., Sheth, A., and Das, A.
\newblock The brittleness of ai-generated image watermarking techniques: Examining their robustness against visual paraphrasing attacks.
\newblock \emph{arXiv preprint arXiv:2408.10446}, 2024.

\bibitem[Bauer et~al.(2024)Bauer, Trapp, Stenger, Leppich, Kounev, Leznik, Chard, and Foster]{bauer2024comprehensiveexplorationsyntheticdata}
Bauer, A., Trapp, S., Stenger, M., Leppich, R., Kounev, S., Leznik, M., Chard, K., and Foster, I.
\newblock Comprehensive exploration of synthetic data generation: A survey, 2024.

\bibitem[Bhardwaj et~al.(2021)Bhardwaj, Kelleher, Costabello, and O'Sullivan]{bhardwaj2021adversarial}
Bhardwaj, P., Kelleher, J., Costabello, L., and O'Sullivan, D.
\newblock Adversarial attacks on knowledge graph embeddings via instance attribution methods.
\newblock \emph{arXiv preprint arXiv:2111.03120}, 2021.

\bibitem[Bojchevski \& G{\"u}nnemann(2019)Bojchevski and G{\"u}nnemann]{bojchevski2019adversarial}
Bojchevski, A. and G{\"u}nnemann, S.
\newblock Adversarial attacks on node embeddings via graph poisoning.
\newblock In \emph{International conference on machine learning}, pp.\  695--704. PMLR, 2019.

\bibitem[Bouritsas et~al.(2023)Bouritsas, Frasca, Zafeiriou, and Bronstein]{9721082}
Bouritsas, G., Frasca, F., Zafeiriou, S., and Bronstein, M.~M.
\newblock Improving graph neural network expressivity via subgraph isomorphism counting.
\newblock \emph{IEEE Transactions on Pattern Analysis and Machine Intelligence}, 45\penalty0 (1):\penalty0 657--668, 2023.

\bibitem[Busbridge et~al.(2019)Busbridge, Sherburn, Cavallo, and Hammerla]{busbridge2019relationalgraphattentionnetworks}
Busbridge, D., Sherburn, D., Cavallo, P., and Hammerla, N.~Y.
\newblock Relational graph attention networks, 2019.

\bibitem[{\c{C}}ano et~al.(2017){\c{C}}ano, Morisio, et~al.]{ccano2017music}
{\c{C}}ano, E., Morisio, M., et~al.
\newblock Music mood dataset creation based on last. fm tags.
\newblock In \emph{2017 International Conference on Artificial Intelligence and Applications, Vienna, Austria}, pp.\  15--26, 2017.

\bibitem[Cao \& Kipf(2022)Cao and Kipf]{decao2022molganimplicitgenerativemodel}
Cao, N.~D. and Kipf, T.
\newblock Molgan: An implicit generative model for small molecular graphs, 2022.

\bibitem[Chang et~al.(2023)Chang, Koulieris, and Shum]{chang2023designfundamentalsdiffusionmodels}
Chang, Z., Koulieris, G.~A., and Shum, H. P.~H.
\newblock On the design fundamentals of diffusion models: A survey, 2023.

\bibitem[Charpenay \& Schockaert(2025)Charpenay and Schockaert]{10.24963/ijcai.2024/364}
Charpenay, V. and Schockaert, S.
\newblock Capturing knowledge graphs and rules with octagon embeddings.
\newblock In \emph{Proceedings of the Thirty-Third International Joint Conference on Artificial Intelligence}, 2025.

\bibitem[Chen \& Wornell(2001)Chen and Wornell]{chen2001class}
Chen, B. and Wornell, G.
\newblock A class of provably good methods for digital watermarking and information embedding.
\newblock \emph{IEEE Trans. on Information Theory}, 47\penalty0 (4):\penalty0 291--314, 2001.

\bibitem[Chen et~al.(2024)Chen, Fang, Cai, Huang, and Sun]{10.5555/3666122.3667344}
Chen, S., Fang, H., Cai, Y., Huang, X., and Sun, M.
\newblock Differentiable neuro-symbolic reasoning on large-scale knowledge graphs.
\newblock In \emph{Proceedings of the 37th International Conference on Neural Information Processing Systems}, Red Hook, NY, USA, 2024. Curran Associates Inc.

\bibitem[Chen et~al.(2019)Chen, Huang, Xu, Guo, Guo, Sun, Li, Pfadler, Zhao, and Zhao]{10.1145/3292500.3330652}
Chen, W., Huang, P., Xu, J., Guo, X., Guo, C., Sun, F., Li, C., Pfadler, A., Zhao, H., and Zhao, B.
\newblock Pog: Personalized outfit generation for fashion recommendation at alibaba ifashion.
\newblock In \emph{Proceedings of the 25th ACM SIGKDD International Conference on Knowledge Discovery \& Data Mining}, KDD '19, pp.\  2662–2670. Association for Computing Machinery, 2019.
\newblock ISBN 9781450362016.
\newblock \doi{10.1145/3292500.3330652}.

\bibitem[Cox et~al.(2007)Cox, Miller, Bloom, Fridrich, and Kalker]{cox2007digital}
Cox, I., Miller, M., Bloom, J., Fridrich, J., and Kalker, T.
\newblock \emph{Digital watermarking and steganography}.
\newblock Morgan kaufmann, 2007.

\bibitem[Das et~al.(2024)Das, Kong, Sen, and Zhou]{10.5555/3692070.3692474}
Das, A., Kong, W., Sen, R., and Zhou, Y.
\newblock A decoder-only foundation model for time-series forecasting.
\newblock In \emph{Proceedings of the 41st International Conference on Machine Learning}. JMLR.org, 2024.

\bibitem[Dathathri et~al.(2024)Dathathri, See, Ghaisas, Huang, McAdam, Welbl, Bachani, Kaskasoli, Stanforth, Matejovicova, Hayes, Vyas, Merey, Brown-Cohen, Bunel, Balle, Cemgil, Ahmed, Stacpoole, Shumailov, Baetu, Gowal, Hassabis, and Kohli]{Dathathri2024}
Dathathri, S., See, A., Ghaisas, S., Huang, P.-S., McAdam, R., Welbl, J., Bachani, V., Kaskasoli, A., Stanforth, R., Matejovicova, T., Hayes, J., Vyas, N., Merey, M.~A., Brown-Cohen, J., Bunel, R., Balle, B., Cemgil, T., Ahmed, Z., Stacpoole, K., Shumailov, I., Baetu, C., Gowal, S., Hassabis, D., and Kohli, P.
\newblock Scalable watermarking for identifying large language model outputs.
\newblock \emph{Nature}, 634\penalty0 (8035):\penalty0 818--823, 2024.

\bibitem[Dong et~al.(2022)Dong, Wang, Wang, Derr, and Li]{10.1145/3534678.3539319}
Dong, Y., Wang, S., Wang, Y., Derr, T., and Li, J.
\newblock On structural explanation of bias in graph neural networks.
\newblock In \emph{Proceedings of the 28th ACM SIGKDD Conference on Knowledge Discovery and Data Mining}, pp.\  316–326, New York, NY, USA, 2022. Association for Computing Machinery.

\bibitem[Fan et~al.(2019)Fan, Ma, Yin, Wang, Tang, and Li]{10.1145/3298689.3347011}
Fan, W., Ma, Y., Yin, D., Wang, J., Tang, J., and Li, Q.
\newblock Deep social collaborative filtering.
\newblock In \emph{Proceedings of the 13th ACM Conference on Recommender Systems}, pp.\  305–313, New York, NY, USA, 2019. Association for Computing Machinery.

\bibitem[Fridrich et~al.(2002)Fridrich, Goljan, and Du]{fridrich2002lossless}
Fridrich, J., Goljan, M., and Du, R.
\newblock Lossless data embedding—new paradigm in digital watermarking.
\newblock \emph{EURASIP Journal on Advances in Signal Processing}, 2002:\penalty0 1--12, 2002.

\bibitem[Glasserman(2004)]{glasserman2004monte}
Glasserman, P.
\newblock Monte carlo methods in financial engineering, 2004.

\bibitem[Han et~al.(2025)Han, Wang, Shomer, Guo, Ding, Lei, Halappanavar, Rossi, Mukherjee, Tang, He, Hua, Long, Zhao, Shah, Javari, Xia, and Tang]{han2025retrievalaugmentedgenerationgraphsgraphrag}
Han, H., Wang, Y., Shomer, H., Guo, K., Ding, J., Lei, Y., Halappanavar, M., Rossi, R.~A., Mukherjee, S., Tang, X., He, Q., Hua, Z., Long, B., Zhao, T., Shah, N., Javari, A., Xia, Y., and Tang, J.
\newblock Retrieval-augmented generation with graphs (graphrag), 2025.

\bibitem[Hu et~al.(2024)Hu, Jiang, Guo, and Gong]{hu2024transfer}
Hu, Y., Jiang, Z., Guo, M., and Gong, N.
\newblock A transfer attack to image watermarks.
\newblock \emph{arXiv preprint arXiv:2403.15365}, 2024.

\bibitem[Ji et~al.(2022)Ji, Pan, Cambria, Marttinen, and Yu]{Ji_2022}
Ji, S., Pan, S., Cambria, E., Marttinen, P., and Yu, P.~S.
\newblock A survey on knowledge graphs: Representation, acquisition, and applications.
\newblock \emph{IEEE Transactions on Neural Networks and Learning Systems}, 33\penalty0 (2):\penalty0 494–514, 2022.

\bibitem[Jiang et~al.(2024)Jiang, Cao, Xiao, Bhatia, Sun, and Han]{jiang2024kgfit}
Jiang, P., Cao, L., Xiao, C., Bhatia, P., Sun, J., and Han, J.
\newblock {KG}-{FIT}: Knowledge graph fine-tuning upon open-world knowledge.
\newblock In \emph{The Thirty-eighth Annual Conference on Neural Information Processing Systems}, 2024.

\bibitem[Kirchenbauer et~al.(2023)Kirchenbauer, Geiping, Wen, Katz, Miers, and Goldstein]{kirchenbauer2023watermark}
Kirchenbauer, J., Geiping, J., Wen, Y., Katz, J., Miers, I., and Goldstein, T.
\newblock A watermark for large language models.
\newblock In \emph{International Conference on Machine Learning}, pp.\  17061--17084. PMLR, 2023.

\bibitem[Le et~al.(2024)Le, Zhong, Liu, Xu, Chaudhary, Zhou, and Xu]{pmlr-v235-le24c}
Le, D., Zhong, S., Liu, Z., Xu, S., Chaudhary, V., Zhou, K., and Xu, Z.
\newblock Knowledge graphs can be learned with just intersection features.
\newblock In Salakhutdinov, R., Kolter, Z., Heller, K., Weller, A., Oliver, N., Scarlett, J., and Berkenkamp, F. (eds.), \emph{Proceedings of the 41st International Conference on Machine Learning}, pp.\  26199--26214. PMLR, 2024.

\bibitem[LIANG et~al.(2024)LIANG, Liu, Li, Meng, Liu, Wang, sihang zhou, and Liu]{liang2024clustering}
LIANG, K., Liu, Y., Li, H., Meng, L., Liu, S., Wang, S., sihang zhou, and Liu, X.
\newblock Clustering then propagation: Select better anchors for knowledge graph embedding.
\newblock In \emph{The Thirty-eighth Annual Conference on Neural Information Processing Systems}, 2024.

\bibitem[Liu et~al.(2024)Liu, Pan, Lu, Li, Hu, Zhang, Wen, King, Xiong, and Yu]{liu2024surveytextwatermarkingera}
Liu, A., Pan, L., Lu, Y., Li, J., Hu, X., Zhang, X., Wen, L., King, I., Xiong, H., and Yu, P.~S.
\newblock A survey of text watermarking in the era of large language models, 2024.

\bibitem[Liu et~al.(2025)Liu, Ke, Wang, Wang, Gao, Shang, Li, Xu, Ji, and Li]{10.24963/ijcai.2024/243}
Liu, J., Ke, W., Wang, P., Wang, J., Gao, J., Shang, Z., Li, G., Xu, Z., Ji, K., and Li, Y.
\newblock Fast and continual knowledge graph embedding via incremental lora.
\newblock In \emph{Proceedings of the Thirty-Third International Joint Conference on Artificial Intelligence}, 2025.

\bibitem[Nandi et~al.(2024)Nandi, Kaur, Singla, and Mausam]{Nandi2023DynaSembleDE}
Nandi, A., Kaur, N., Singla, P., and Mausam.
\newblock Dynasemble: Dynamic ensembling of textual and structure-based models for knowledge graph completion.
\newblock In \emph{Annual Meeting of the Association for Computational Linguistics}, 2024.

\bibitem[Patnaik(1949)]{patnaik1949non}
Patnaik, P.
\newblock The non-central $\chi$ 2-and f-distribution and their applications.
\newblock \emph{Biometrika}, 36\penalty0 (1/2):\penalty0 202--232, 1949.

\bibitem[Podell et~al.(2023)Podell, English, Lacey, Blattmann, Dockhorn, Müller, Penna, and Rombach]{podell2023sdxlimprovinglatentdiffusion}
Podell, D., English, Z., Lacey, K., Blattmann, A., Dockhorn, T., Müller, J., Penna, J., and Rombach, R.
\newblock Sdxl: Improving latent diffusion models for high-resolution image synthesis, 2023.

\bibitem[Radford et~al.(2021)Radford, Kim, Hallacy, Ramesh, Goh, Agarwal, Sastry, Askell, Mishkin, Clark, Krueger, and Sutskever]{DBLP:conf/icml/RadfordKHRGASAM21}
Radford, A., Kim, J.~W., Hallacy, C., Ramesh, A., Goh, G., Agarwal, S., Sastry, G., Askell, A., Mishkin, P., Clark, J., Krueger, G., and Sutskever, I.
\newblock Learning transferable visual models from natural language supervision.
\newblock In Meila, M. and Zhang, T. (eds.), \emph{Proceedings of the 38th International Conference on Machine Learning, {ICML} 2021, 18-24 July 2021, Virtual Event}, pp.\  8748--8763. {PMLR}, 2021.

\bibitem[Rombach et~al.(2021)Rombach, Blattmann, Lorenz, Esser, and Ommer]{rombach2021highresolution}
Rombach, R., Blattmann, A., Lorenz, D., Esser, P., and Ommer, B.
\newblock High-resolution image synthesis with latent diffusion models, 2021.

\bibitem[Schlichtkrull et~al.(2017)Schlichtkrull, Kipf, Bloem, van~den Berg, Titov, and Welling]{schlichtkrull2017modelingrelationaldatagraph}
Schlichtkrull, M., Kipf, T.~N., Bloem, P., van~den Berg, R., Titov, I., and Welling, M.
\newblock Modeling relational data with graph convolutional networks, 2017.

\bibitem[Shomer et~al.(2023)Shomer, Jin, Wang, and Tang]{10.1145/3543507.3583544}
Shomer, H., Jin, W., Wang, W., and Tang, J.
\newblock Toward degree bias in embedding-based knowledge graph completion.
\newblock In \emph{Proceedings of the ACM Web Conference 2023}, pp.\  705–715, New York, NY, USA, 2023. Association for Computing Machinery.

\bibitem[Simonovsky \& Komodakis(2018)Simonovsky and Komodakis]{10.1007/978-3-030-01418-6_41}
Simonovsky, M. and Komodakis, N.
\newblock Graphvae: Towards generation of small graphs using variational autoencoders.
\newblock In K{\r{u}}rkov{\'a}, V., Manolopoulos, Y., Hammer, B., Iliadis, L., and Maglogiannis, I. (eds.), \emph{Artificial Neural Networks and Machine Learning -- ICANN 2018}, pp.\  412--422, Cham, 2018. Springer International Publishing.

\bibitem[Smits \& Borghuis(2022)Smits and Borghuis]{Smits2022}
Smits, J. and Borghuis, T.
\newblock \emph{Generative AI and Intellectual Property Rights}, pp.\  323--344.
\newblock T.M.C. Asser Press, The Hague, 2022.

\bibitem[Song et~al.(2022)Song, Meng, and Ermon]{song2022denoisingdiffusionimplicitmodels}
Song, J., Meng, C., and Ermon, S.
\newblock Denoising diffusion implicit models, 2022.

\bibitem[Sun et~al.(2019)Sun, Deng, Nie, and Tang]{sun2019rotate}
Sun, Z., Deng, Z.-H., Nie, J.-Y., and Tang, J.
\newblock Rotate: Knowledge graph embedding by relational rotation in complex space.
\newblock \emph{arXiv preprint arXiv:1902.10197}, 2019.

\bibitem[Veličković et~al.(2018)Veličković, Cucurull, Casanova, Romero, Liò, and Bengio]{veličković2018graphattentionnetworks}
Veličković, P., Cucurull, G., Casanova, A., Romero, A., Liò, P., and Bengio, Y.
\newblock Graph attention networks, 2018.

\bibitem[Vero et~al.(2024)Vero, Balunovi\'{c}, and Vechev]{10.5555/3692070.3694089}
Vero, M., Balunovi\'{c}, M., and Vechev, M.
\newblock Cuts: customizable tabular synthetic data generation.
\newblock In \emph{Proceedings of the 41st International Conference on Machine Learning}. JMLR.org, 2024.

\bibitem[Vignac et~al.(2023)Vignac, Krawczuk, Siraudin, Wang, Cevher, and Frossard]{vignac2023digressdiscretedenoisingdiffusion}
Vignac, C., Krawczuk, I., Siraudin, A., Wang, B., Cevher, V., and Frossard, P.
\newblock Digress: Discrete denoising diffusion for graph generation, 2023.

\bibitem[Wang et~al.(2018)Wang, Zhang, Xie, and Guo]{wang2018dkn}
Wang, H., Zhang, F., Xie, X., and Guo, M.
\newblock Dkn: Deep knowledge-aware network for news recommendation.
\newblock In \emph{Proceedings of the 2018 world wide web conference}, pp.\  1835--1844, 2018.

\bibitem[Wang et~al.(2023)Wang, Liang, Li, Li, Ghanem, Zimmermann, Zhou, Yi, Zhang, and Wang]{wang2023brave}
Wang, K., Liang, Y., Li, X., Li, G., Ghanem, B., Zimmermann, R., Zhou, Z., Yi, H., Zhang, Y., and Wang, Y.
\newblock Brave the wind and the waves: Discovering robust and generalizable graph lottery tickets.
\newblock \emph{IEEE Transactions on Pattern Analysis and Machine Intelligence}, 46\penalty0 (5):\penalty0 3388--3405, 2023.

\bibitem[Wang et~al.(2025)Wang, Zhang, Zhou, Wu, Yu, Zhao, Yin, Fu, Yan, Luo, et~al.]{wang2025comprehensive}
Wang, K., Zhang, G., Zhou, Z., Wu, J., Yu, M., Zhao, S., Yin, C., Fu, J., Yan, Y., Luo, H., et~al.
\newblock A comprehensive survey in llm (-agent) full stack safety: Data, training and deployment.
\newblock \emph{arXiv preprint arXiv:2504.15585}, 2025.

\bibitem[Wang et~al.(2019)Wang, He, Wang, Feng, and Chua]{10.1145/3331184.3331267}
Wang, X., He, X., Wang, M., Feng, F., and Chua, T.-S.
\newblock Neural graph collaborative filtering.
\newblock In \emph{Proceedings of the 42nd International ACM SIGIR Conference on Research and Development in Information Retrieval}, pp.\  165–174, New York, NY, USA, 2019. Association for Computing Machinery.

\bibitem[Wen et~al.(2023)Wen, Kirchenbauer, Geiping, and Goldstein]{NEURIPS2023_b54d1757}
Wen, Y., Kirchenbauer, J., Geiping, J., and Goldstein, T.
\newblock Tree-rings watermarks: Invisible fingerprints for diffusion images.
\newblock In Oh, A., Naumann, T., Globerson, A., Saenko, K., Hardt, M., and Levine, S. (eds.), \emph{Advances in Neural Information Processing Systems}, pp.\  58047--58063. Curran Associates, Inc., 2023.

\bibitem[Wen et~al.(2024)Wen, Kirchenbauer, Geiping, and Goldstein]{wen2024tree}
Wen, Y., Kirchenbauer, J., Geiping, J., and Goldstein, T.
\newblock Tree-rings watermarks: Invisible fingerprints for diffusion images.
\newblock \emph{Advances in Neural Information Processing Systems}, 36, 2024.

\bibitem[Wu et~al.(2020)Wu, Qiao, Chen, Wu, Qi, Lian, Liu, Xie, Gao, Wu, and Zhou]{wu-etal-2020-mind}
Wu, F., Qiao, Y., Chen, J.-H., Wu, C., Qi, T., Lian, J., Liu, D., Xie, X., Gao, J., Wu, W., and Zhou, M.
\newblock {MIND}: A large-scale dataset for news recommendation.
\newblock In Jurafsky, D., Chai, J., Schluter, N., and Tetreault, J. (eds.), \emph{Proceedings of the 58th Annual Meeting of the Association for Computational Linguistics}, pp.\  3597--3606, Online, July 2020. Association for Computational Linguistics.
\newblock \doi{10.18653/v1/2020.acl-main.331}.

\bibitem[Yan \& Han(2002)Yan and Han]{yan2002gspan}
Yan, X. and Han, J.
\newblock gspan: Graph-based substructure pattern mining.
\newblock In \emph{2002 IEEE International Conference on Data Mining, 2002. Proceedings.}, pp.\  721--724. IEEE, 2002.

\bibitem[Yang(2024)]{yang2024guisegraphgaussianshading}
Yang, R.
\newblock Guise: Graph gaussian shading watermark, 2024.

\bibitem[Yang et~al.(2024{\natexlab{a}})Yang, Chen, and Xiang]{Yang2024}
Yang, Y., Chen, J., and Xiang, Y.
\newblock A review on the reliability of knowledge graph: from a knowledge representation learning perspective.
\newblock \emph{World Wide Web}, 28\penalty0 (1):\penalty0 4, 2024{\natexlab{a}}.

\bibitem[Yang et~al.(2024{\natexlab{b}})Yang, Zeng, Chen, Fang, Zhang, and Yu]{Yang_2024_CVPR}
Yang, Z., Zeng, K., Chen, K., Fang, H., Zhang, W., and Yu, N.
\newblock Gaussian shading: Provable performance-lossless image watermarking for diffusion models.
\newblock In \emph{Proceedings of the IEEE/CVF Conference on Computer Vision and Pattern Recognition (CVPR)}, pp.\  12162--12171, 2024{\natexlab{b}}.

\bibitem[Yin et~al.(2024)Yin, Wang, and Song]{yin2024rethinking}
Yin, H., Wang, Z., and Song, Y.
\newblock Rethinking complex queries on knowledge graphs with neural link predictors.
\newblock In \emph{The Twelfth International Conference on Learning Representations}, 2024.

\bibitem[You et~al.(2018)You, Ying, Ren, Hamilton, and Leskovec]{pmlr-v80-you18a}
You, J., Ying, R., Ren, X., Hamilton, W., and Leskovec, J.
\newblock {G}raph{RNN}: Generating realistic graphs with deep auto-regressive models.
\newblock In Dy, J. and Krause, A. (eds.), \emph{Proceedings of the 35th International Conference on Machine Learning}, pp.\  5708--5717. PMLR, 2018.

\bibitem[Yu et~al.(2021)Yu, Skripniuk, Abdelnabi, and Fritz]{yu2021artificial}
Yu, N., Skripniuk, V., Abdelnabi, S., and Fritz, M.
\newblock Artificial fingerprinting for generative models: Rooting deepfake attribution in training data.
\newblock In \emph{Proceedings of the IEEE/CVF International conference on computer vision}, pp.\  14448--14457, 2021.

\bibitem[Yu et~al.(2023)Yu, Zhang, and Deng]{YU2023110534}
Yu, Z., Zhang, C., and Deng, C.
\newblock An improved gnn using dynamic graph embedding mechanism: A novel end-to-end framework for rolling bearing fault diagnosis under variable working conditions.
\newblock \emph{Mechanical Systems and Signal Processing}, 200:\penalty0 110534, 2023.

\bibitem[Zhang et~al.(2019)Zhang, Zheng, Gao, Miao, Su, Li, and Ren]{ijcai2019p674}
Zhang, H., Zheng, T., Gao, J., Miao, C., Su, L., Li, Y., and Ren, K.
\newblock Data poisoning attack against knowledge graph embedding.
\newblock In \emph{Proceedings of the Twenty-Eighth International Joint Conference on Artificial Intelligence, {IJCAI-19}}, pp.\  4853--4859. International Joint Conferences on Artificial Intelligence Organization, 2019.

\bibitem[Zhang et~al.(2024{\natexlab{a}})Zhang, Liu, i~Martin, Bearfield, Brun, and Guan]{zhang2024attackresilient}
Zhang, L., Liu, X., i~Martin, A.~V., Bearfield, C.~X., Brun, Y., and Guan, H.
\newblock Attack-resilient image watermarking using stable diffusion.
\newblock In \emph{The Thirty-eighth Annual Conference on Neural Information Processing Systems}, 2024{\natexlab{a}}.

\bibitem[Zhang et~al.(2024{\natexlab{b}})Zhang, Li, Yu, Xu, Li, and Zhang]{zhang2024editguard}
Zhang, X., Li, R., Yu, J., Xu, Y., Li, W., and Zhang, J.
\newblock Editguard: Versatile image watermarking for tamper localization and copyright protection.
\newblock In \emph{Proceedings of the IEEE/CVF Conference on Computer Vision and Pattern Recognition}, pp.\  11964--11974, 2024{\natexlab{b}}.

\bibitem[Zhao et~al.(2023{\natexlab{a}})Zhao, Ananth, Li, and Wang]{zhao2023provablerobustwatermarkingaigenerated}
Zhao, X., Ananth, P., Li, L., and Wang, Y.-X.
\newblock Provable robust watermarking for ai-generated text, 2023{\natexlab{a}}.

\bibitem[Zhao et~al.(2023{\natexlab{b}})Zhao, Zhang, Su, Vasan, Grishchenko, Kruegel, Vigna, Wang, and Li]{zhao2023invisible}
Zhao, X., Zhang, K., Su, Z., Vasan, S., Grishchenko, I., Kruegel, C., Vigna, G., Wang, Y.-X., and Li, L.
\newblock Invisible image watermarks are provably removable using generative ai.
\newblock \emph{arXiv preprint arXiv:2306.01953}, 2023{\natexlab{b}}.

\bibitem[Zhao et~al.(2023{\natexlab{c}})Zhao, Pang, Du, Yang, Cheung, and Lin]{zhao2023recipe}
Zhao, Y., Pang, T., Du, C., Yang, X., Cheung, N.-M., and Lin, M.
\newblock A recipe for watermarking diffusion models.
\newblock \emph{arXiv preprint arXiv:2303.10137}, 2023{\natexlab{c}}.

\bibitem[Zhou et~al.(2024)Zhou, Zhang, Yao, quanming yao, and Han]{zhou2024less}
Zhou, Z., Zhang, Y., Yao, J., quanming yao, and Han, B.
\newblock Less is more: One-shot subgraph reasoning on large-scale knowledge graphs.
\newblock In \emph{The Twelfth International Conference on Learning Representations}, 2024.

\bibitem[Zhu et~al.(2024{\natexlab{a}})Zhu, Chen, Yan, Huang, Lin, Li, Tu, Hu, Hu, and Wang]{zhu2024genimage}
Zhu, M., Chen, H., Yan, Q., Huang, X., Lin, G., Li, W., Tu, Z., Hu, H., Hu, J., and Wang, Y.
\newblock Genimage: A million-scale benchmark for detecting ai-generated image.
\newblock \emph{Advances in Neural Information Processing Systems}, 36, 2024{\natexlab{a}}.

\bibitem[Zhu et~al.(2024{\natexlab{b}})Zhu, Takahashi, and Kataoka]{Zhu2024WatermarkembeddedAE}
Zhu, P., Takahashi, T., and Kataoka, H.
\newblock Watermark-embedded adversarial examples for copyright protection against diffusion models.
\newblock \emph{2024 IEEE/CVF Conference on Computer Vision and Pattern Recognition (CVPR)}, pp.\  24420--24430, 2024{\natexlab{b}}.

\end{thebibliography}
\bibliographystyle{icml2025}

\newpage
\appendix
\onecolumn

\section{Defending Against Isomorphism and Structural Variations} \label{appendix Normalization}

\begin{algorithm}[H]
\caption{Graph Alignment}
\label{alg:graph_normalization}
\begin{algorithmic}[1]
\REQUIRE Input graph $G = (V, E)$ with vertex attributes and adjacency matrix $A$
\ENSURE Normalized graph $\hat{G}$

\STATE For each $v \in V$: 
\STATE \quad $d[v] \gets \deg(v)$, $c[v] \gets c(v)$  \quad \text{Compute the degree and clustering coefficient for each vertex}

\STATE $V_o \gets \text{Sort}(V, d, c)$ \quad \text{Sort $V$ first by degree $d[v]$ and then by clustering coefficient $c[v]$ in ascending order}

\STATE $A' \gets \text{Reorder}(A, V_o)$ \quad \text{Reorder rows and columns of $A$ according to the order in $V_o$}

\STATE $\hat{G} \gets \emptyset$ 

\FOR{$v_i \in V_o$}
    \STATE $\hat{G} \gets \hat{G} \cup \{v_i\}$ \quad \text{Add vertex $v_i$ to $\hat{G}$}
\ENDFOR

\FOR{$(v_i, v_j) \in E$}
    \STATE $\hat{G} \gets \hat{G} \cup \{(v_i, v_j)\}$ \quad \text{Add edge $(v_i, v_j)$ to $\hat{G}$}
\ENDFOR

\FOR{$v_i \in V_o$}
    \STATE $\text{RearrangeAttributes}(v_i, \hat{G})$ \quad \text{Rearrange or update attributes of $v_i$ based on the new order in $\hat{G}$}
\ENDFOR

\RETURN $\hat{G}$
\end{algorithmic}
\end{algorithm}

\textbf{Symbol Definitions:}
\begin{itemize}
    \item $d[v]$: The degree of vertex $v$, i.e., the number of edges incident to $v$.
    \item $c[v]$: The clustering coefficient of vertex $v$ measures the degree to which vertices in a graph tend to cluster together.
    \item $\hat{G}$: The normalized graph after vertex and edge processing.
    \item $\text{RearrangeAttributes}(v_i, \hat{G})$: A function to reorder or update the attributes of vertex $v_i$ in the normalized graph $\hat{G}$ based on the sorted vertex list $V_o$.
\end{itemize}

Algorithm~\ref{alg:graph_normalization} outlines a graph alignment process designed to enhance the robustness of a graph against isomorphism and structural variations, which are common forms of attack. The process begins by calculating the degree and clustering coefficient for each node. These attributes reflect the node's connectivity and the density of its neighbors, which are crucial for determining the node's importance within the graph.

Next, the nodes are sorted based on these attributes, first by degree and then by clustering coefficient. This sorting ensures that nodes with lower connectivity are processed first, while more central nodes appear later. Following the sorting, the graph's adjacency matrix is reordered to match the new node order, preserving the graph’s original relationships.

A new graph, $\hat{G}$, is created by adding nodes in the sorted order. As each node is added, its attributes are updated according to the new structure, ensuring consistency. Finally, the edges are added to the new graph based on the original connections, but aligned with the new node ordering.

The output is a normalized graph that is less susceptible to structural perturbations. This process helps improve the graph's resilience to attacks that alter its structure, making it more stable for applications like watermarking, where the goal is to embed and extract watermarks while maintaining robustness against manipulation.

\begin{algorithm}
\caption{Redundant Embedding Based on Subgraphs}
\label{alg:Redundant}
\begin{algorithmic}[1]
\STATE \textbf{Input:} Graph $G = (V, E)$, Smallest embedded size $s$, Watermark $W$
\STATE \textbf{Output:} Watermarked graph $G'$
\STATE $l \gets \lfloor |V| / s \rfloor$ \quad \text{Compute the number of communities}

\STATE $C \gets \{C_1, C_2, \dots, C_l\}$ 
\STATE when $\bigcup_{i=1}^l C_i = V$ and $C_i \cap C_j = \emptyset, \forall i \neq j$ \quad\text{Partition $G$ into $l$ non-overlapping communities}

\FOR{each community $c \in C$}
    \STATE $V_{C_i} \gets \{v \mid v \in C_i, \quad\text{selected based on a predefined strategy}\}$
    \STATE $c_{norm} \gets$ GraphNormalization($c$, $A_c$) \quad\text{Normalize the graph structure}
    \FOR{each vertex $v \in V_c$}
        \STATE EmbedWatermark($v$, $W$) \quad\text{Embed watermark into selected vertices}
    \ENDFOR
\ENDFOR

\STATE \RETURN $G'$

\end{algorithmic}
\end{algorithm}

Algorithm~\ref{alg:Redundant} describes a process for redundantly embedding a watermark into a graph based on its subgraphs. The aim is to enhance the robustness of the watermark against structural modifications and attacks by embedding it in multiple subgraph communities.

The process starts by computing the number of communities $l$ based on the graph's total number of vertices $|V|$ and a predefined smallest embedded size $s$. The graph $G$ is then partitioned into $l$ non-overlapping communities $C_1, C_2, \dots, C_l$, ensuring that every vertex in the graph is assigned to exactly one community, and the communities do not overlap.

Next, for each community $c \in C$, the algorithm selects a subset of vertices, $V_{C_i}$, based on a predefined strategy, which could be influenced by factors such as node importance or connectivity. The subgraph corresponding to each community, $c$, is then normalized using the `GraphNormalization` function. This step ensures that the internal structure of each community is standardized, making the watermark embedding more robust to structural variations.

After normalizing the graph structure, the watermark is embedded into selected vertices of each community using the `EmbedWatermark` function. This embedding process is repeated for each vertex in every community. By embedding the watermark redundantly across different communities of the graph, the method increases the likelihood of watermark retention even if parts of the graph undergo modification or attack.

Finally, the watermarked graph $G'$ is returned, which contains the embedded watermark in a redundant and robust manner, ensuring better resilience to structural changes or adversarial perturbations. This approach is particularly useful for applications that require high security, such as digital watermarking in graphs, where the goal is to protect the integrity of the watermark despite potential attacks.

\section{Case Study in Embedding Space} \label{appendix Case Study}

\begin{figure*}[ht]
\centering

\begin{subfigure}[b]{0.49\textwidth}
    \centering
    \includegraphics[width=\textwidth]{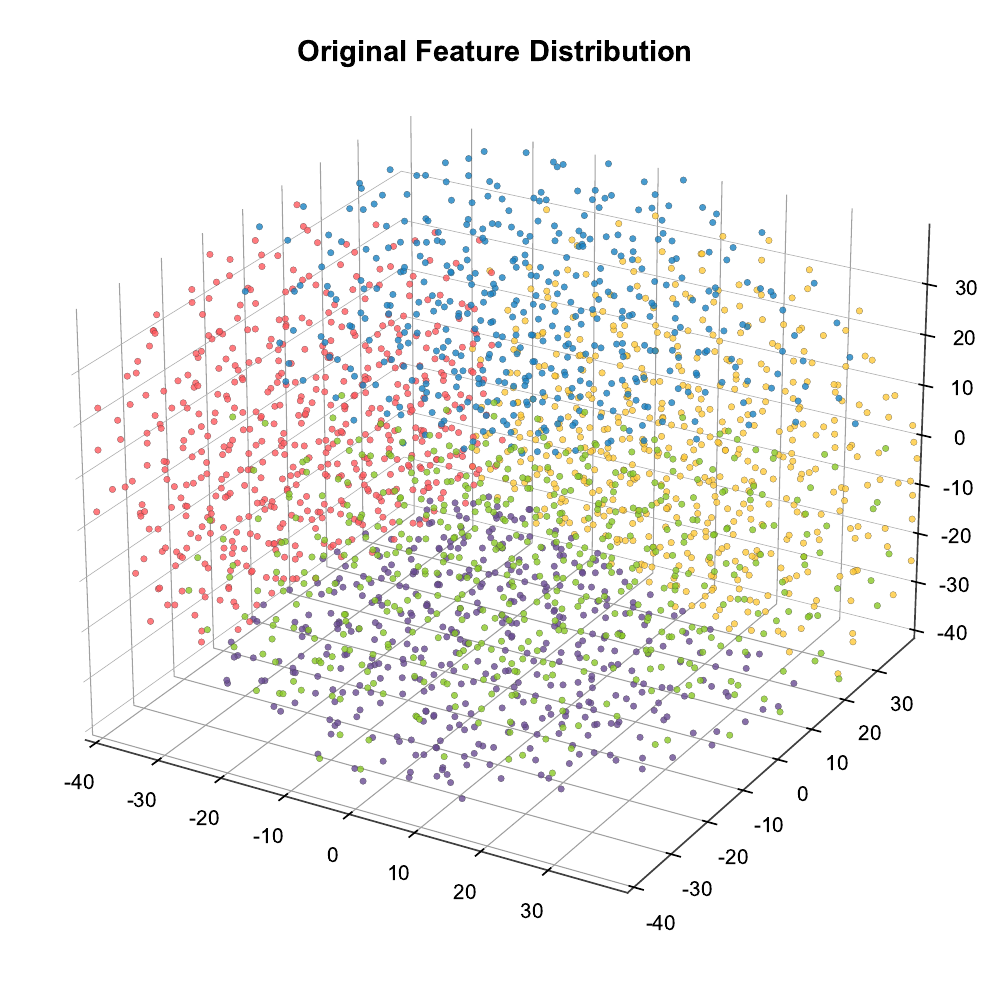}
    \label{fig:density_fixed}
\end{subfigure}
\hfill
\begin{subfigure}[b]{0.49\textwidth}
    \centering
    \includegraphics[width=\textwidth]{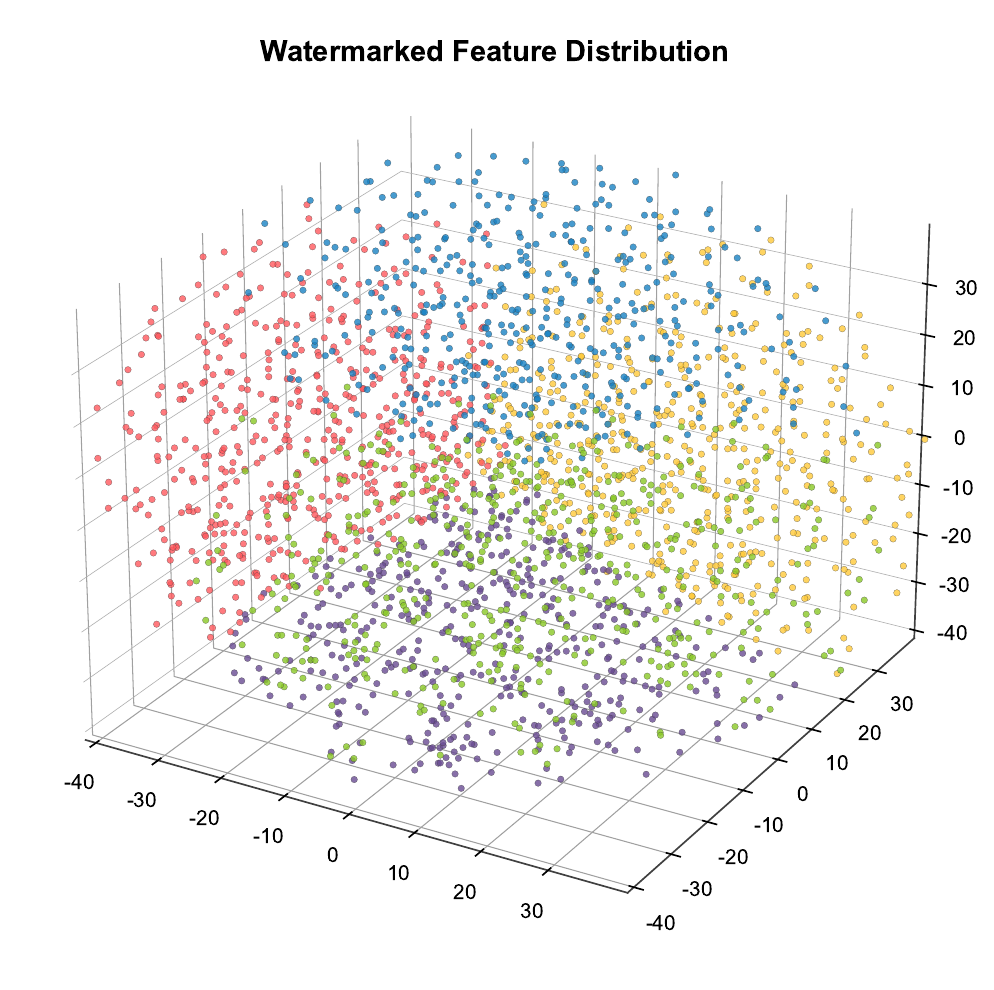}
    \label{fig:density_lawmm}
\end{subfigure}

\caption{Ablation study of the density of the watermark mask matrix on all datasets. Each subfigure corresponds to a different method.}
\label{fig:apdx_case}
\end{figure*}
To further demonstrate the effectiveness of our proposed watermarking method, a case study is conducted on the dataset AliF related to customer behavior analysis. We select 2400 entity embeddings from five communities (A-E) from AliF for watermark embedding and visualize the embedding vectors before and after processing by dimensionality reduction. The case background is presented in Table~\ref{tab: casebk}, the result of the visualization is shown in Figure~\ref{fig:apdx_case}

\begin{table*}[ht]
\centering
\caption{Case Background.}
\resizebox{\textwidth}{!}{%
\begin{tabular}{ccccc|cccc|c}
\toprule
\multicolumn{5}{c|}{\textbf{Number of Entities}} & \multirow{2}{*}{\textbf{total}} & \multirow{2}{*}{\textbf{Dimensions of Embeddings}}  &  \multirow{2}{*}{\textbf{DDIM Inference Steps}} & \multirow{2}{*}{\textbf{Density of Mask}} & \multirow{2}{*}{\textbf{Cosine Similarity}} \\
\cmidrule(lr){1-5} 
 A & B & C & D & E & & && &\\
\midrule
524 & 507 & 524 & 420 & 425& 2400 & 4096 & 75 & 0.015 & 0.9535  \\
\bottomrule
\end{tabular}
}
\label{tab: casebk}
\end{table*}

For visualization purposes, we used the t-SNE technique to reduce the dimensionality of the embedding vectors from 4096 to 3. This technique is particularly suitable for mapping data from high-dimensional to two-dimensional or three-dimensional space while preserving local structural relationships between data points.

The similarity between points $x_i$ and $x_j$ in a high-dimensional space is defined by the conditional probability $p_{j|i}$ as follows, where $\|x_i - x_j\|^2$ represents the squared Euclidean distance. $\sigma_i$ is a parameter that controls the width of the neighborhood of the point $x_i$.
\begin{equation}
p_{j|i} = \frac{\exp\left(-\|x_i - x_j\|^2 / 2\sigma_i^2 \right)}{\sum_{k \neq i} \exp\left(-\|x_i - x_k\|^2 / 2\sigma_i^2 \right)}
\end{equation}

To symmetrize the similarity, the joint probability is defined as follows, where $n$ is the total number of data points.
\begin{equation}
p_{ij} = \frac{p_{j|i} + p_{i|j}}{2n}
\end{equation}

In a low-dimensional space, the similarity between the points $y_i$ and $y_j$ is defined by the t-distribution as:
\begin{equation}
q_{ij} = \frac{\left(1 + \|y_i - y_j\|^2 \right)^{-1}}{\sum_{k \neq l} \left(1 + \|y_k - y_l\|^2 \right)^{-1}}
\end{equation}

t-SNE uses Kullback-Leibler (KL) divergence as a dissimilarity measure between a high-dimensional distribution $p_{ij}$and a low-dimensional distribution $q_{ij}$. The goal is to minimize the following loss function:
\begin{equation}
\mathcal{L} = \sum_{i \neq j} p_{ij} \log \frac{p_{ij}}{q_{ij}}
\end{equation}

t-SNE optimizes the loss function $\mathcal{L}$ using Stochastic Gradient Descent (SGD). During the optimization, the position of the low-dimensional embedding $y_i$ is gradually adjusted, so that the low-dimensional similarity $q_{ij}$ is closer to the high-dimensional similarity $p_{ij}$.

In Figure~\ref{fig:apdx_case}, it can be found that the color gradient and position distribution of the points in the figure almost remain unchanged before and after the embedding of the watermark, indicating that the embedding of the watermark does not significantly change the geometric structure of the original embedding. In addition, both images show a spherical distribution that gradually diffuses from the center to the outside, and the density of the points remains the same, indicating that the characteristics of the original vector embedding are well preserved after embedding the watermark. The watermark embedding may introduce only slight perturbations in some local details or some dimensions, but these perturbations are not sufficient to cause significant effects on the overall distribution. 

\begin{table*}[ht]
\centering
\caption{Case Result.}
\resizebox{0.5\textwidth}{!}{%
\begin{tabular}{ccc}
\toprule
\textbf{Community} & \textbf{P-value} & \textbf{Is Detected}  \\
\midrule
A & $8.29e-27$ & \ding{51}   \\
B & $1.52e-26$ & \ding{51}   \\
C & $4.20e-19$ & \ding{51}   \\
D & $1.88e-15$ & \ding{51}   \\
E & $3.27e-34$ & \ding{51}   \\
\bottomrule
\end{tabular}
}
\label{tab: case}
\end{table*}
Then, we carry out watermark detection on the watermarked knowledge graph, and we select the significance level as 5e-5. We report their P-values to check whether the watermark is detected.
In summary, our watermarking method performs well in terms of transparency. It neither destroys nor enhances the distribution characteristics of the original embeddings, but also achieves certain robustness and detectability.



\section{Latent Diffusion Model} \label{appendix DDIM}

\subsection{Variational Autoencoder}
To effectively capture the intricate structure of graph data, we employ a Variational Autoencoder (VAE). This advanced model maps the entity embeddings of the knowledge graph into a latent space, where the underlying patterns and relationships can be more efficiently represented and analyzed. To ensure that we can adequately and accurately capture the complex patterns inherent in the graph structure, we utilize multiple Relational Graph Attention (RGAT) modules. These modules are specifically designed to handle the relational nature of the graph data, allowing the model to focus on different types of relationships and entities within the graph. By leveraging the power of multiple RGAT modules, we can produce higher-quality latent representations that better encapsulate the essential characteristics of the graph structure. These embeddings are then sampled using the reparameterization trick. This technique enables us to introduce stochasticity in a differentiable manner, which is crucial for the training process of the VAE. Finally, the sampled embeddings are decoded back into the reconstructed entity embeddings, allowing us to evaluate the quality of the learned representations and the model's overall performance in capturing and reconstructing the graph data.

\subsection{Diffusion, DDIM and Inversion} \label{appendix: DDIM}

Denoising Diffusion Implicit Models (DDIM) provide an efficient alternative to standard diffusion-based generative models by leveraging a non-Markovian forward and reverse process. This section introduces the DDIM sampling process, its inversion, and the necessary mathematical formulation and derivation.

\textbf{Forward Diffusion Process: }

The forward process in DDIM maps a data point $\mathbf{Z}_0 \in \mathbb{R}^d$ to a noisy latent variable $\mathbf{Z}_T$ over $T$ timesteps. It is defined as:
\begin{equation}
\mathbf{Z}_t = \sqrt{\alpha_t} \mathbf{Z}_0 + \sqrt{1 - \alpha_t} \mathbf{\epsilon}, \quad \mathbf{\epsilon} \sim \mathcal{N}(\mathbf{0}, \mathbf{I}), \label{eq:forward}
\end{equation}
where $\alpha_t$ is a scheduling parameter that controls the noise variance at timestep $t$.

\textbf{Reverse Diffusion Process: }

The reverse process reconstructs $\mathbf{Z}_0$ from $\mathbf{Z}_T$ by iteratively denoising the latent variable. DDIM modifies the reverse process to make it deterministic, defined as:
\begin{equation}
\mathbf{Z}_{t-1} = \sqrt{\alpha_{t-1}} \mathbf{\hat{Z}}_0 + \sqrt{1 - \alpha_{t-1}} \mathbf{\epsilon}_t, \label{eq:reverse}
\end{equation}
where $\mathbf{\hat{Z}}_0$ is the predicted clean data at timestep $t$ and is obtained via:
\begin{equation}
\mathbf{\hat{Z}}_0 = \frac{\mathbf{Z}_t - \sqrt{1 - \alpha_t} \mathbf{\epsilon}_t}{\sqrt{\alpha_t}}. \label{eq:x0_pred}
\end{equation}

The noise $\mathbf{\epsilon}_t$ is predicted using a pre-trained noise estimation model, typically parameterized as $\epsilon_\theta(\mathbf{Z}_t, t)$.

\textbf{DDIM Sampling: }

DDIM employs a deterministic sampling strategy by reparameterizing the reverse process. Specifically, the update step becomes:
\begin{equation}
\mathbf{Z}_{t-1} = \sqrt{\alpha_{t-1}} \mathbf{\hat{Z}}_0 + \sqrt{1 - \alpha_{t-1}} \cdot \eta \cdot \mathbf{\epsilon}_t, \label{eq:ddim_sampling}
\end{equation}
where $\eta \in [0, 1]$ controls the stochasticity of the process. Setting $\eta = 0$ results in a fully deterministic process, while $\eta > 0$ introduces controlled randomness.

\textbf{Inversion Process: }

DDIM supports exact inversion, enabling the recovery of $\mathbf{Z}_t$ from $\mathbf{Z}_{t-1}$. By rearranging Eq.~\eqref{eq:ddim_sampling}, the inversion is expressed as:
\begin{equation}
\mathbf{Z}_t = \sqrt{\alpha_t} \mathbf{\hat{Z}}_0 + \sqrt{1 - \alpha_t} \cdot \eta \cdot \mathbf{\epsilon}_t. \label{eq:ddim_inversion}
\end{equation}
This exact inversion capability is crucial for tasks such as embedding watermarks or debugging generative processes.

\section{Watermark Embedding in Frequency Domain} \label{appendix FD}

\begin{figure*}[ht]
\centering
\includegraphics[width=1.0\textwidth]{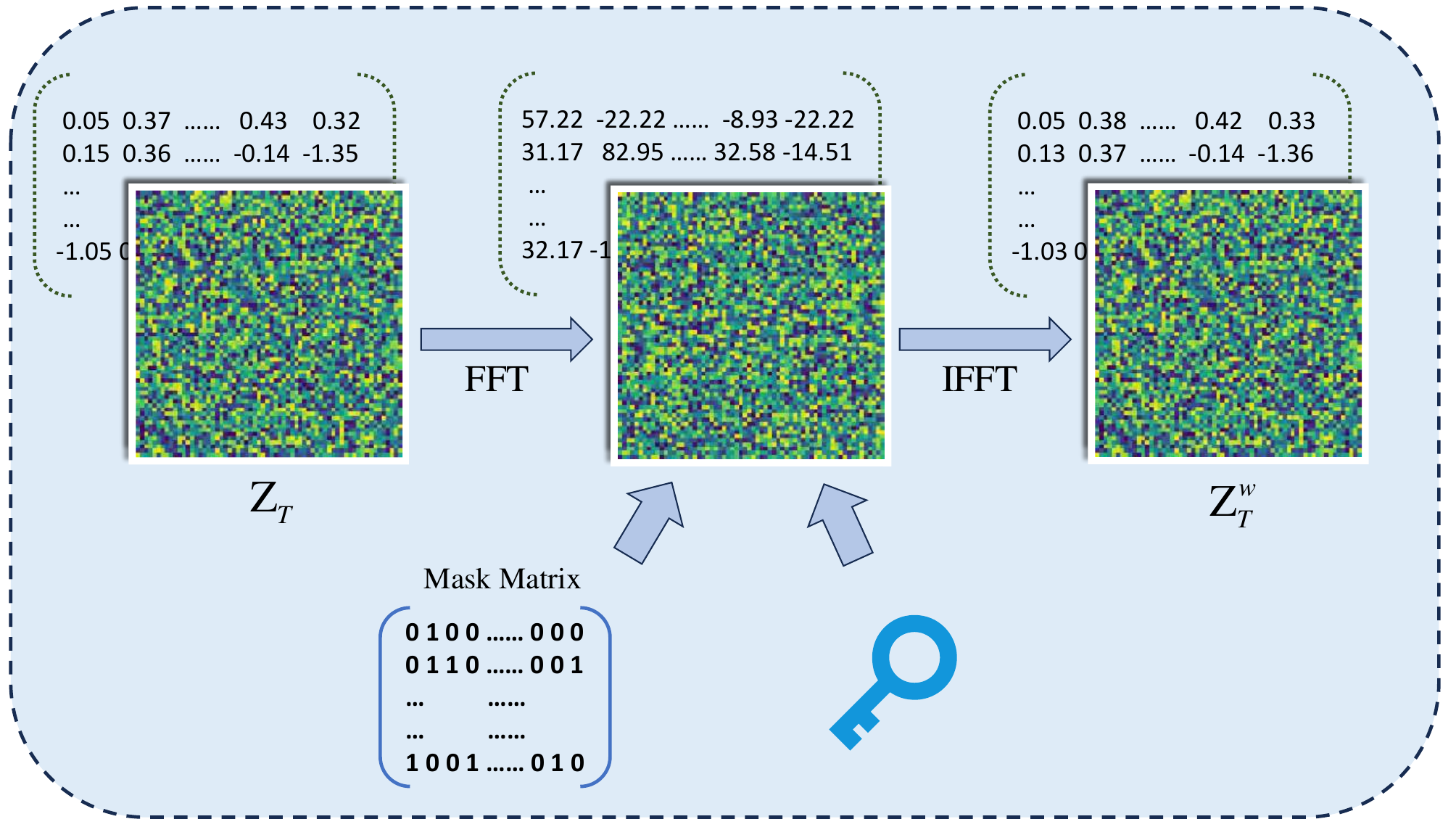}
\caption{Watermark Embedding in Frequency Domain.}
\label{fig: FFT}
\end{figure*}

The theoretical basis of watermark embedding in the frequency domain is mainly based on the frequency domain transformation technology in signal processing. The essence of the Fourier transform we employ is to separate the time domain and frequency domain characteristics, and the Fourier transform decomposes the time domain signal into frequency domain components; that is, the signal is represented as a superposition of sine waves with different frequencies. In the frequency domain, Low-frequency components usually contain the main energy of the signal (such as the overall structure or low-resolution features). High-frequency components: Describe rapid changes in the signal (such as detail or texture information). The watermark is embedded in some specific frequency components of the frequency domain (such as middle frequency or high frequency), which can not only hide the watermark but also reduce the impact on the data's original structure. The frequency domain properties after the Fourier transform are robust to certain linear operations such as compression, smoothing, rotation, and cropping: for example, the rotation operation only causes a phase change in the frequency domain, while the amplitude spectrum remains stable. This makes the method of embedding the watermark in the frequency domain highly robust to these transformations. 

Our method uses the energy distribution characteristics, perceptual sensitivity, and frequency domain stability of the Fourier transform frequency domain to embed the watermark into the frequency domain, and then realizes the flexible and adaptive selection of the frequency region through our LAWMM to achieve the goal of invisibility, robustness, and detectability of the embedded watermark. These theoretical foundations provide a solid foundation for frequency domain watermarking technology and also explain its advantages in the face of common attacks such as compression, geometric transformation, and smoothing.

\section{Metrics} \label{appendix MEC}

\subsection{Detectability and Robustness}
We use the AUC as an authoritative indicator for evaluating the detectability and robustness of our watermarking methods.
\begin{itemize}
    \item \textbf{Area Under Curve (AUC)}: 
    AUC refers to the area under the Receiver Operating Characteristic (ROC) curve, a widely used metric for evaluating classification models. The AUC score ranges from 0 to 1, with a value closer to 1 indicating better model performance. It represents the probability that a randomly chosen positive sample will be ranked higher than a randomly chosen negative sample. Mathematically, AUC can be expressed as:
\begin{equation}
\text{AUC} = \frac{1}{N_p N_n} \sum_{i=1}^{N_p} \sum_{j=1}^{N_n} \mathbb{I}(y_i > y_j)
\end{equation}
where \( N_p \) and \( N_n \) are the numbers of positive and negative samples, respectively, \( y_i \) and \( y_j \) are the predicted scores of the positive and negative samples, and \( \mathbb{I}(y_i > y_j) \) is the indicator function that is 1 if \( y_i > y_j \), and 0 otherwise.

\end{itemize}
\subsection{Transparency}
We evaluate the transparency of watermarking from two dimensions: the \textbf{similarity} of knowledge graph embedding before and after watermarking and the \textbf{quality} of watermarked knowledge graph. Therefore, we used the following two categories of evaluation metrics.

We use cosine similarity to evaluate the \textbf{similarity} of knowledge graphs.
\begin{itemize}
    \item \textbf{Cosine Similarity}: Cosine similarity is a similarity measure between two non-zero vectors in an inner product space. It is widely used in many fields, including information retrieval, natural language processing, and machine learning. The cosine similarity between two vectors \( \mathbf{A} \) and \( \mathbf{B} \) is defined as:
\begin{equation}
\text{cosine similarity}(\mathbf{A}, \mathbf{B}) = \frac{\mathbf{A} \cdot \mathbf{B}}{\|\mathbf{A}\| \|\mathbf{B}\|}
\end{equation}
\end{itemize}

For quality, we use the following four metrics. Given a set of test triples $\mathcal{T} = \{(h, r, t)\}$ in a knowledge graph $\mathcal{G}$, let $r(h, t)$ denote the rank of the correct entity $t$ (or $h$) among all candidates in a ranking task.
\begin{itemize}
    \item \textbf{Geometric Mean Rank (GMR)}: GMR captures the overall ranking tendency in a multiplicative rather than additive manner, making it particularly suitable for datasets where ranking distributions exhibit heavy tails. It provides insights into the overall ranking distribution, making it useful for comprehensive model evaluation. The GMR is defined as follows:
\begin{equation}
    \text{GMR} = \left( \prod_{(h,r,t) \in \mathcal{T}} r(h,t) \right)^{\frac{1}{|\mathcal{T}|}}.
\end{equation}

    \item \textbf{Harmonic Mean Rank (HMR)}: HMR mitigates the influence of extremely high ranks, ensuring that poor predictions do not disproportionately impact the evaluation. Moreover, HMR is expressed directly in rank space, making it more interpretable while still benefiting from reciprocal weighting. 
\begin{equation}
    \text{HMR} = \frac{|\mathcal{T}|}{\sum_{(h,r,t) \in \mathcal{T}} \frac{1}{r(h,t)}}.
\end{equation}

    \item  \textbf{Arithmetic Mean Rank (AMR)}: AMR directly reflects the average rank of correct entities, making it easy to understand and compare across models. A major limitation of AMR is that it is heavily influenced by extremely high-rank values, which can distort performance evaluation.
\begin{equation}
    \text{AMR} = \frac{1}{|\mathcal{T}|} \sum_{(h,r,t) \in \mathcal{T}} r(h, t).
\end{equation}

    \item \textbf{Hits@k}: Hits@$k$ provides a direct measure of retrieval success within a fixed rank cutoff, making it useful for real-world applications requiring top-$k$ recommendations.
\begin{equation}
    \text{Hits@}k = \frac{1}{|\mathcal{T}|} \sum_{(h,r,t) \in \mathcal{T}} \mathbb{I}[r(h,t) \leq k],
\end{equation}
where $\mathbb{I}[\cdot]$ is the indicator function that returns 1 if the condition holds and 0 otherwise. This metric evaluates whether the correct entity appears within the top-$k$ ranks.

\end{itemize}



\section{Notations and Definitions}

\begin{table}[ht]
\centering
\caption{Notations and Definitions}
\begin{tabular}{lp{12cm}}
\Xhline{1.2pt}
\textbf{Notation} & \textbf{Definition} \\ 
\Xhline{1.2pt}

\( F(\cdot) \) & Fourier Transform \\

\( M \) & Watermark mask matrix \\

\( S \) & Watermark signature \\

\( \sigma^2 \) & Variance of the watermark signature \\

\( Z^w_{T} \) & Watermarked latent representation \\

\( Z^{INV}_T \) & Initial noise vector recovered via DDIM inversion \\

\( L \) & Loss function in watermark embedding \\

\( \epsilon \) & Threshold for latent space equilibrium \\

\( \delta \) & Total perturbation in an attack \\

\( \delta_k \) & Perturbation applied to each subgraph \( G_k \) \\

\( q \) & Norm type in the attack objective \\

\( Y \) & Fourier transform result during watermark extraction \\

\( \mu_i \) & Mean of \( Y \) \\

\( \sigma_i^2 \) & Variance of \( Y \) \\

\( T \) & Test statistic for watermark detection \\

\( \lambda \) & Likelihood ratio test statistic \\

\( \hat{T} \) & Simplified test statistic \\

\( p \) & P-value for watermark detection \\

\( \alpha \) & Significance level for watermark detection \\

\( \rho \) & Density of the watermark mask matrix \\

\( \Phi(C_i) \) & Watermark embedding in the community layer \\

\( \Psi(v) \) & Watermark embedding in the vertex layer \\

\( C(W) \) & Effective information capacity of watermark encoding \\

\( H(\cdot) \) & Entropy \\

\( D(G, \tilde{G}) \) & Divergence metric between graphs \( G \) and \( \tilde{G} \) \\

\( L^\dagger \) & Pseudoinverse of the graph Laplacian \\

\( \Delta A \) & Perturbation matrix for the graph \\

\( \eta(v) \) & Centrality of vertex \( v \) \\

\( \hat{Z}_0 \) & Predicted clean data in DDIM inversion \\

\( \eta \) & Stochasticity control parameter in DDIM sampling \\

\( \epsilon_t \) & Predicted noise in DDIM reverse process \\

\( \alpha_t \) & Scheduling parameter in DDIM forward diffusion process \\

\( \sigma_i \) & Parameter controlling neighborhood width in t-SNE \\

\( p_{ij} \) & Joint probability in high-dimensional space \\

\( q_{ij} \) & t-distribution similarity in low-dimensional space \\

\( L \) & Loss function in t-SNE optimization \\

\Xhline{1.2pt}
\end{tabular}
\label{tab:your_table_label}
\end{table}

\end{document}